\documentclass[12pt]{JHEP3}

\def\S{{\cal S}}
\def\H{{\cal H}}
\def\N{{\cal N}}

\def\bs{{\bf h}}

\def\be{\begin{equation}}
\def\ee{\end{equation}}
\def\bea{\begin{eqnarray}}
\def\eea{\end{eqnarray}}

\newcommand{\su}{{{ hu(1,1|(0,1,2)\!\!:\!\![M,2]) }}}
\newcommand{\hc}{{{ hc(1|2\!\!:\!\![M,2]) }}}
\newcommand{\hco}{{{ hc(1|(1,2)\!\!:\!\![M,2]) }}}
\newcommand{\hu}{{{ hu(1|2\!\!:\!\![M,2]) }}}
\newcommand{\hupm}{{{ hu^E_\pm(1|(1,2)\!\!:\!\![M,2]) }}}
\newcommand{\huo}{{{ hu(1|(1,2)\!\!:\!\![M,2]) }}}
\newcommand{\hue}{{{ hu^E(1|(1,2)\!\!:\!\![M,2]) }}}
\newcommand{\hunmpq}{{{ hu(n,m|(u,v,2)\!\!:\!\![M,2]) }}}
\newcommand{\hunm}{{{ hu(n,m|(0,1,2)\!\!:\!\![M,2]) }}}
\newcommand{\honm}{{{ ho(n,m|(0,1,2)\!\!:\!\![M,2]) }}}
\newcommand{\huspnm}{{{ husp(n,m|(0,1,2)\!\!:\!\![M,2]) }}}
\newcommand{\honmpm}{{{ ho_\pm(n,m|(0,1,2)\!\!:\!\![M,2]) }}}
\newcommand{\huspnmpm}{{{ husp_\pm (n,m,|(0,1,2)\!\!:\!\![M,2]) }}}
\newcommand{\hunmpm}{{{ hu_\pm(n,m|(0,1,2)\!\!:\!\![M,2]) }}}

\newcommand{\supm}{{{ hu_\pm(1,1|(0,1,2)\!\!:\!\![M,2]) }}}

\newsavebox{\ver}
\newsavebox{\gor}
\newsavebox{\verp}
\newsavebox{\gorp}
\newsavebox{\toch}

\newcommand{\bee}{\begin{eqnarray}}
\newcommand{\eee}{\end{eqnarray}}
\newcommand{\nn}{\nonumber}

\newcommand{\go}{\omega}

\newcommand{\f}{\frac}
\newcommand{\p}{\partial}
\newcommand{\half}{\frac{1}{2}}
\newcommand{\ga}{\alpha}

\newcommand{\bga}{{\bar{\alpha}}}
\newcommand{\ba}{{\bar{a}}}
\newcommand{\bgb}{{\bar{\beta}}}

\newcommand{\gb}{\beta}
\newcommand{\gga}{\gamma}

\newcommand{\gep}{\epsilon}
\newcommand{\lla}{\langle \langle}
\newcommand{\rr}{\rangle \rangle}

\newcommand{\gvep}{\varepsilon}
\newcommand{\gs}{\sigma}

\newcommand\un{{\underline{n}}}

\newcommand\ls{\!\!\!\!\!\!\!}

\title{
 Higher Spin Superalgebras in any Dimension\\
and their Representations}

\bigskip
\author
{M.A.~Vasiliev\\
  {\small
  \it
I.E.Tamm Department of Theoretical Physics, Lebedev Physical Institute,\\
Leninsky prospect 53, 119991, Moscow, Russia}
}

\abstract { Fock module realization for the unitary singleton
representations of the $d-1$ dimensional conformal algebra
$o(d-1,2)$, which correspond to the spaces of single-particle states
of massless scalar and spinor in $d-1$ dimensions, is given. The
pattern of the tensor product of a pair of singletons is analyzed in
any dimension. It is shown that for $d>3$ the tensor product of two
boson singletons decomposes into a sum of all integer spin totally
symmetric massless representations in $AdS_d$, the tensor product of
boson and fermion singletons gives a sum of all half-integer spin
symmetric massless representations in $AdS_d$, and the tensor
product of two fermion singletons in $d>4$ gives rise to massless
fields of mixed symmetry types in $AdS_d$ depicted by Young tableaux
with one row and one column together with certain totally
antisymmetric massive fields. In the special case of $o(2,2)$,
tensor products of $2d$ massless scalar and/or spinor modules
contain infinite sets of $2d$ massless conformal fields of different
spins. The obtained  results extend the $4d$ result of Flato and
Fronsdal  \cite{FF} to any dimension and provide a nontrivial
consistency check for the recently proposed higher spin model in
$AdS_d$ \cite{d}. We define a class of higher spin superalgebras
which act on the supersingleton and  higher spin states in any
dimension.  For the cases of $AdS_3$, $AdS_4$, and $AdS_5$ the
isomorphisms with the higher spin superalgebras defined earlier in
terms of spinor generating elements are established. }

\begin{document}

\maketitle

\def\theequation{\arabic{section}.\arabic{equation}}
\setcounter{equation}0



\section{Introduction}
In the paper \cite{d} nonlinear equations of motion for
interacting totally symmetric
massless bosonic fields of all spins in any dimension
have been formulated. The primary goal of this paper is to
show that the global higher spin (HS) symmetry algebras of \cite{d}
admit massless unitary representations which correspond to the sets
of massless fields of the models of \cite{d}.
This  provides a nontrivial consistency check of the results of
\cite{d} analogous to that carried out in \cite{KV,KV1} for the
$4d$ HS models.

One of the key results of this paper consists of the extension to
any dimension of the theorem of Flato and Fronsdal \cite{FF} which
states that the tensor products of pairs of $AdS_4$ singletons give
rise to sums of all $AdS_4$ massless representations of $o(3,2)\sim
sp(4)$. The generalization of the Flato-Fronsdal theorem for the
$AdS_5$ case, which is of most interest from the superstring theory
side, was obtained in \cite{dou} in terms of doubletons (see also
\cite{FZ}). Analogous analysis of the case of $AdS_7$ was performed
in \cite{SS7}. The key element for these constructions was the
oscillator realization of the space-time symmetry algebras and their
superextensions based on the low-dimensional isomorphisms like
$o(2,2)\sim sp(2)\oplus sp(2)$, $o(3,2) \sim sp(4)$ and  $o(4,2)\sim
su(2,2)$ which allow realizations of space-time superalgebras in
terms of bilinears of oscillators carrying spinor representations of
space-time symmetry (super)algebras. The singleton and doubleton
representations in lower dimensions admit a simple realization of
Fock modules associated with the these spinor oscillators (see
\cite{Gun1} and references therein). In these terms the
Flato-Fronsdal theorem can be proved directly by decomposing the
tensor product of two such Fock modules into irreducible submodules
of the same symmetry algebra (see for example \cite{KV1} for the
$AdS_4$ case). However, the realization in terms of spinors does not
work beyond some lower dimensions because the afore mentioned
isomorphisms do not take place for general $d$. Nevertheless, as we
show, the analysis can be performed in any dimension within the
realization of the orthogonal algebra $o(M,2)$ in terms of bosonic
oscillators carrying  $o(M,2)$ vector indices. In this case the
corresponding  Fock module also plays the key role. The difference
between the two constructions is that this Fock module forms a
reducible $o(M,2)$--module and, to single out the unitary singleton
submodules, some additional restrictions on the carrier space have
to be imposed. A somewhat unusual feature is that the corresponding
submodules do not contain the Fock vacuum. Otherwise, the extension
of the Flato-Fronsdal theorem to any dimension is quite uniform.

The Flato-Fronsdal theorem and its higher
dimensional extensions provide a
group-theoretical basis for
the $AdS/CFT$ correspondence conjecture
\cite{AdSCFT,GKP,W1} and are especially important for the analysis
of the correspondence between $d$ dimensional
boundary conformal models and HS gauge models in the bulk
$AdS_{d+1}$.  The latter issue was  addressed in a number of
papers in different contexts \cite{AdSHS}-\cite{Schnitzer:2003zr}.
(A closely related issue is the analysis of the tensionless limit of
string in $AdS$; see,  for example,
\cite{Dhar:2003fi}-\cite{Bonelli:2003zu}.)  Our
group-theoretical analysis agrees with the results of
\cite{Das:2003vw,Gopakumar:2003ns},
where the sector of a boundary conformal scalar field
in any dimension was discussed, and suggests the extension of these
results to the models with boundary and bulk fermions in any
dimension.

The bosonic HS algebra of \cite{d} is the conformal HS algebra of a
massless boundary scalar \cite{East} in $d-1$ dimensions, i.e. it is
the infinite dimensional symmetry algebra of the massless
Klein-Gordon equation. Another class of HS algebras is associated
with the massless boundary spinors. In the case of $d=4$ these two
algebras were isomorphic.  As we show this is not true beyond $d=4$.
The  $AdS_d$ bulk gauge fields corresponding to bilinears of
massless boundary scalar form the set of all totally symmetric
massless bosonic fields. The $AdS_d$ bulk gauge fields of the
conformal HS algebras corresponding to bilinears of a boundary
spinor are bosons having mixed symmetry described by  Young tableaux
with one row and one column of various lengths and heights,
respectively. For odd $d$ (i.e., even-dimensional boundary theory)
there are generically three sorts of HS algebras: nonchiral type $A$
algebras and two type $B$ chiral algebras which correspond to chiral
boundary spinors. In the type $B$ cases antisymmetric HS bulk
tensors satisfy certain (anti)selfduality conditions.

Not surprisingly, the scalar$\times$scalar and spinor$\times$spinor
HS algebras are two bosonic subalgebras of some
HS superalgebra in $AdS_d$ with any $d$. The fermionic sector
of the corresponding bulk $AdS_d$ gauge fields consists of all totally
symmetric half-integer spin fields in $AdS_d$. We will argue that
all constructed (super)algebras  underly some
consistent HS gauge theories in $AdS_d$.
It is important to note that the infinite dimensional HS superalgebras
constructed in this paper contain finite dimensional SUSY
 subalgebras only for some lower dimensions that admit
equivalent description in terms of spinor twistor variables.

The content of this paper is as follows.
In the rest of the Introduction we summarize some relevant facts
on the unitary representations of the $AdS_d$ algebra
$o(d-1,2)$ (subsection \ref{Anti-de Sitter algebra}) and
discuss some general properties of the HS algebras (subsection
\ref{General conditions on higher spin algebras}) focusing main
attention on the admissibility condition which gives
a criterion that allows one to single out those algebras which
can be symmetries of a consistent field-theoretical
model. In section \ref{Simplest bosonic higher spin algebras}
we define simplest bosonic HS algebras. The projection technics
useful for the analysis of quotient algebras is introduced
in section \ref{Projection technics}. The HS superalgebras are
defined in section \ref{Higher Spin Superalgebra}. In section
\ref{Spinorial realization} it is shown that for the particular case
of $AdS_4$ the HS superalgebra admits equivalent realization
of \cite{Fort2,FVA} in terms of spinors, and analogous construction is
discussed for the $AdS_3$ and $AdS_5$ HS algebras.
 Oscillator (Fock) realization for the unitary
representation of single-particle states of the boundary conformal
scalar and spinor are constructed in sections \ref{rac} and
\ref{di}, respectively. In section \ref{FF}, the pattern of the
tensor products of these modules is found and it is shown that HS
superalgebras discussed in this paper satisfy the admissibility
condition. Unfolded formulation of the free field equations for
boundary conformal fields is briefly discussed in section
\ref{Unfolded equations for conformal fields}. Section
\ref{Conclusion} contains conclusions. In Appendix we collect some
useful facts on the description of Young tableaux in terms of
oscillators.

\subsection{Anti-de Sitter algebra}
\label{Anti-de Sitter algebra}
HS algebras are specific infinite dimensional extensions
of one or another
$d$ dimensional space-time symmetry (super)algebra $g$.
In this paper we will be mainly interested in the
$AdS_d$ case of  $g=o(d-1,2)$.
The generators $T^{AB}$ of $o(M,2)$
satisfy the commutation relations
\be
[T^{AB}\,,T^{CD}]= \eta^{BC} T^{AD} - \eta^{AC} T^{BD}
-\eta^{BD} T^{AC} +\eta^{AD} T^{BD}\,,
\ee
where  $\eta^{AB}$ is the invariant symmetric  form of
$o(M,2)$ ($A,B = 0, \ldots , M+1$). We will use the mostly minus
convention with $\eta^{00} = \eta^{M+1 M+1}=1$ and
$\eta^{ab} = -\delta^{ab}$ for the space-like values of
$A=a=1\ldots M$. The $AdS_{M+1}$ energy operator is
\be
\label{ener}
E = i T^{M+1\,0}\,.
\ee
The noncompact  generators of $o(M,2)$ are
\be
T^{\pm a} =iT^{0a}\mp  T^{M+1\,a}\,,
\ee
\be
\label{ET}
[E, T^{\pm a}] = \pm T^{\pm a}\,, \qquad
[T^{- a} , T^{+ b} ] = 2(\delta^{ab} E + T^{ab} )\,.
\ee
The compact generators $T^{ab}$ of $o(M)$ commute with
$E$. The generators  $T^{AB}$ are anti-Hermitian,
$(T^{AB})^\dagger = - T^{AB}$,
and, therefore,
\be
\label{herm2}
E^\dagger = E\,,\qquad
(T^{\pm a})^\dagger = T^{\mp a}\,,\qquad (T^{ab})^\dagger = - T^{ab}\,.
\ee

An irreducible bounded energy unitary representation $\H(E_0 , \bs
)$ of $o(M,2)$ is characterized by some eigenvalue $E_0$ of $E$ and
weight $\bs$ of $o(M)$ which refer to the lowest energy (vacuum)
states $|E_0, \bs\rangle$ of $\H(E_0 , \bs )$  that satisfy
$T^{-a} |E_0, \bs \rangle=0$ and form a finite dimensional
module of $o(M)\oplus o(2) \subset o(M,2)$. A value of the
quadratic Casimir operator $C_2 = -\half T^{AB}T_{AB}$ on $\H(E_0
, \bs )$ is \be \label{cas} C_2 = E_0 (E_0 -M) +\gga (\bs)\,, \ee
where $\gga(\bs)$ is the value of the Casimir operator $\gga_2 =
-\half T^{ab}T_{ab}$ of $o(M)$ on the vacuum $|E_0, \bs \rangle$.

As shown by Metsaev \cite{metb}, a bosonic massless field in $AdS_{M+1}$,
which carries ``spin'' corresponding to the representation of $o(M)$
with the weights $\bs$, has the vacuum energy
\be
\label{mass}
E^{bos}_0 (\bs) = h^{max} - p-1+M
\ee
where $h^{max}$ is the length of the first row of the
$o(M)$ Young tableau associated with the vacuum space
$|E_0, \bs\rangle$ while $p$ is the number of rows of length
$h^{max}$ at the condition that the total number of rows (i.e., $o(M)$
weights) does not exceed $\half M$ (that can always be achieved
by dualization with the help of the epsilon symbol).
In other words $h^{max}$ and $p$ are, respectively,
the length and height of the upper rectangular block of the
$o(M)$ Young tableau associated with the vacuum weight  $\bs$
(relevant definitions and
facts on Young tableaux are collected in the Appendix),
\sbox{\toch}{\circle*{1}}
\sbox{\gor}{\line(1,0){5}}
\sbox{\ver}{\line(0,1){5}}
\bee   
\begin{picture}(150,110)(0,5)%
{
%
\put(0,0){\line(0,1){40}}
\multiput(3,03)(7,0){04}{\usebox{\toch}}%
\multiput(3,12)(7,0){05}{\usebox{\toch}}%
\multiput(3,21)(7,0){06}{\usebox{\toch}}%
\multiput(0,30)(5,0){9}{\usebox{\gor}}%
\multiput(0,30)(5,0){10}{\usebox{\ver}}%
\multiput(0,35)(5,0){11}{\usebox{\gor}}%
\multiput(0,35)(5,0){12}{\usebox{\ver}}%
\multiput(0,40)(5,0){11}{\usebox{\gor}}%
\multiput(0,40)(5,0){12}{\usebox{\ver}}%
\multiput(0,45)(5,0){13}{\usebox{\gor}}%
\multiput(0,45)(5,0){14}{\usebox{\ver}}%
\multiput(0,50)(5,0){13}{\usebox{\gor}}%
\multiput(0,50)(5,0){14}{\usebox{\ver}}%
\multiput(0,55)(5,0){13}{\usebox{\gor}}%
\multiput(0,55)(5,0){14}{\usebox{\ver}}%
\multiput(0,60)(5,0){13}{\usebox{\gor}}%
\multiput(0,60)(5,0){14}{\usebox{\ver}}%
%
\multiput(0,65)(5,0){19}{\usebox{\gor}}%
\multiput(0,65)(5,0){20}{\usebox{\ver}}%
\multiput(0,70)(5,0){19}{\usebox{\gor}}%
\multiput(0,70)(5,0){20}{\usebox{\ver}}%
\multiput(0,75)(5,0){19}{\usebox{\gor}}%
\multiput(0,75)(5,0){20}{\usebox{\ver}}%
\multiput(0,80)(5,0){19}{\usebox{\gor}}%
\multiput(0,80)(5,0){20}{\usebox{\ver}}%
\multiput(0,85)(5,0){19}{\usebox{\gor}}%
\multiput(0,85)(5,0){20}{\usebox{\ver}}%
\multiput(0,90)(5,0){19}{\usebox{\gor}}%
\multiput(0,90)(5,0){20}{\usebox{\ver}}%
\multiput(0,95)(5,0){19}{\usebox{\gor}}%
\multiput(0,95)(5,0){20}{\usebox{\ver}}%
\multiput(0,100)(5,0){19}{\usebox{\gor}}%
\multiput(0,100)(5,0){20}{\usebox{\ver}}%
\multiput(0,105)(5,0){19}{\usebox{\gor}}%
\put(50,108){$ h^{max}$}
\put(105,90){$p$}
}
\end{picture}
\eee

The expression for lowest energies of fermionic massless
representations is analogous
\cite{metf}
\be
\label{massf}
E^{fer}_0 (\bs ) = h^{max} - p-3/2+M
\ee
where, again, $h^{max}$ and $p$ are, respectively,
the length and height of the upper rectangular block of the
$o(M)$ Young tableau  associated with the $\gamma$-transverse
tensor-spinor realization of the vacuum space
(i.e., $|E_0, \bs\rangle$ is realized
as a space of $o(M)$ tensors  carrying an additional $o(M)$ spinor index,
with the $o(M)$ invariant tracelessness,
 $\gga$-transversality and Young antisymmetry conditions
imposed). A total number of rows of the corresponding Young tableau
does not exceed $\half M$.

More generally, let $D(E_0 ,\bs )$ be a generalized Verma module
induced from some irreducible $o(M)\oplus o(2)$ vacuum module $|E_0,
\bs\rangle$. It is spanned by the states \be \label{hst}
T^{+a_1}\ldots T^{+a_n} |E_0, \bs\rangle\, \ee with various levels
$n$. For the unitary case $D(E_0 ,\bs )= \H(E_0 ,\bs )$ it is
isomorphic to the  Hilbert space of single-particle states of one or
another field-theoretical system, i.e, a space of normalizable
positive-energy solutions of some (irreducible) $o(M,2)$ invariant
field equations in $M+1$ dimensional space-time\footnote{There are
as many independent states (\ref{hst}) as on-mass-shell independent
derivatives of any order of the dynamical fields under consideration
at some point of a $M$ dimensional Cauchy surface.}. Unitarity
implies existence of some invariant positive-definite norm with
respect to which the  Hermiticity conditions (\ref{herm2}) are
satisfied. This requires the vacuum energy $E_0$ to be high enough
\be E_0\geq E_0 (\bs )\,, \ee where $E_0 (\bs )$ is some weight
dependent minimal value of $E_0$ compatible with unitarity. Note
that from the second relation in (\ref{ET}) it follows that $E_0
(\bs ) \geq 0$ in a unitary module.

Starting from inside of the unitarity region and
decreasing $E_0$ for a fixed $\bs$
one approaches the boundary of the
unitarity region, $E_0 = E_0 ({\bs})$.
Some zero-norm vectors then appear in $D (E_0 (\bs) , \bs)$
for $E_0 = E_0 ({\bs})$.
These necessarily should have
vanishing scalar product with any other state
(otherwise there will be
a negative norm state in contradiction with the
assumption that $E_0$ is at the boundary of the unitarity region).
Therefore, the zero-norm states form an invariant subspace called
singular submodule $\S$. By factoring out this subspace one is
left with some unitary module
$\H (E_0 (\bs) , \bs)= D (E_0 (\bs) , \bs) /\S$.
Note that the  submodule $\S$ is induced from
some singular vectors $|E^\prime_0,
\bs^\prime \rangle \in D (E_0 (\bs) , \bs) $
among the states (\ref{hst}) which themselves satisfy the
vacuum condition $T^{-a} |E^\prime_0, \bs^\prime \rangle =0$.

It is well known that the appearance of the null subspace $\S$
manifests gauge symmetries in the underlying field-theoretical
model. More precisely, $\S$ represents leftover on-mass-shell
symmetries with the gauge parameters analogous to the leftover gauge
symmetries of the Maxwell theory $\delta A_n = \p_n \phi$ in the
Lorentz gauge  $\p_n A^n=0\rightarrow \Box \phi =0$. This is because
the space (\ref{hst}) has some fixed values of all Casimir operators
determined by the weights of the vacuum state $|E_0, \bs \rangle$.
As a result, the submodule $\S$ must have the same values of the
Casimir operators. This means in particular that the states of $\S$
satisfy an appropriate Klein-Gordon equation associated with the
quadratic Casimir of $o(M,2)$. As pointed out by Flato and Fronsdal
\cite{ff}, gauge symmetries related with the singular modules can be
of two different types.

Type I is the case of usual gauge symmetry allowing to gauge away
some part of the $AdS_{M+1}$ bulk degrees of freedom of a field
associated with the module $D (E_0 (\bs) , \bs) $ so that the
quotient module $\H (E_0 (\bs) , \bs)= D (E_0 (\bs) , \bs) /\S$
describes a field with local degrees of freedom in $AdS_{M+1}$. The
corresponding fields are gauge fields in $AdS_{M+1}$. We will call
them massless fields as they have minimal lowest energies compatible
with unitarity. Note that the relations (\ref{mass}), (\ref{massf})
for lowest energies of massless fields were derived by Metsaev
\cite{metb,metf} just from the requirements of on-mass-shell gauge
invariance of the corresponding massless equations along with the
unitarity condition\footnote{Let us note that partially massless
fields in $AdS_d$ considered in \cite{DW} correspond to nonunitary
$o(M,2)$-modules resulting from  factorization of submodules of pure
gauge states in the appropriate generalized Verma modules.}. (For
more details on the structure of unitary representations of
noncompact algebras we refer the reader to
 \cite{FEF} and references therein). The massless
representations of $o(M,2)$ of this class are those with  $p <
\f{M}{2}$, i.e. the corresponding vacuum spaces are described by any
$o(M)$--module except for those described by the rectangular Young
tableaux of the maximal height $\half M$ and an arbitrary length.
Note that the latter representations exist only for even $M$ except
for the degenerate cases of tableaux of zero length which correspond
to the lowest energy scalar and spinor $o(M)$--modules for any $M$.

Type II is the case of boundary conformal fields which we will call
singletons when discussing the corresponding unitary
representations. This is the case where all bulk degrees of freedom
are factored out so that the module $\H (E_0 (\bs) , \bs)$ describes
a dynamical system at the boundary of $AdS_{M+1}$. In this case,
$o(M,2)$ acts as conformal group in $M$ dimensions. In accordance
with the results of \cite{sieg, metc} (see also \cite{FEFR}), the
type II representations are those with $p= \f{M}{2}$,  $M$ even, and
minimal energy scalar and spinor $o(M)$ modules, i.e. the
corresponding Young tableau is some rectangular of the maximal
height $\f{M}{2}$ and an arbitrary length (including zero). Indeed,
it is easy to see that these fields form massless representations of
$o(M-1,2)$: dualization of a height $\half M$ tableau with respect
to $o(M-1)$ gives a rectangular block of height $\half M -1$ that
just compensates the effect of replacing $M$ by $M-1$ in
(\ref{mass}) and (\ref{massf}). Also  the appearance of gauge
degrees of freedom in the scalar or spinor modules indicates
decoupling of bulk degrees of freedom \cite{ff}. Note that
field-theoretical realization of this phenomenon was originally
discovered by Dirac \cite{Dir} for the case of $o(3,2)$.

Finally, let us make the following remark. Every lowest weight
unitary $o(M,2)$-module spanned by the vectors (\ref{hst}) forms a
unitary module of $o(M-1,2)\subset o(M,2)$. To find out its
$o(M-1,2)$ pattern one has to decompose a $o(M,2)$--module $D(E_0
,\bs )$ into a direct sum of $o(M-1,2)$--modules. This can be
achieved by looking for vacuum states among
 (\ref{hst}) as those satisfying
$T^{-a^\prime} |E_0^\prime, \bs^\prime \rangle$ =0 with $a^\prime =
1\ldots M-1$. In the $o(M-1)$ covariant basis the states (\ref{hst})
are equivalent to \be \label{hst1} T^{+a^\prime_1}\ldots
T^{+a^\prime_n} (t^+ )^m|E_0, \bs\rangle\,, \qquad t^+ = T^{+M}\,.
\ee Clearly, the dependence on $t^+$  results in the infinite
reducibility of $D(E_0 ,\bs )$ treated as $o(M-1,2)$--module.
 This is expected because infinite
towers of Kaluza-Klein modes should appear. On the
other hand, this tower may be treated as an
infinite dimensional  module of the $M$
dimensional conformal algebra $o(M,2)$ which mixes
fields of different nonzero masses in $AdS_{M}$.

\subsection{General conditions on higher spin algebras}
\label{General conditions on higher spin algebras}

HS algebras are specific infinite dimensional extensions
of one or another
$d$ dimensional space-time symmetry (super)algebra $g$ which in the
$AdS_d$ case is $o(d-1,2)$. Not every extension $h$ of
$g$ gives a HS algebra, however. Here we  summarize
some  general conditions to be satisfied by HS algebras which
may help to rule out  some candidates.

Generally speaking, a HS
algebra is any unbroken global symmetry algebra $h$ of a
vacuum solution of some consistent interacting theory which
contains massless higher spins  in the symmetric vacuum under
consideration. To make it possible to interpret
the model in terms of relativistic fields carrying some masses and spins,
the vacuum solution is demanded to be invariant under
the conventional space-time symmetry $g\subset h$ (e.g, $g=o(M,2)$).
As a global symmetry of a consistent model, $h$  must possess a
unitary representation which contains all massless
states in the model. The same time, any
symmetry parameter of $h$ should me associated with
some gauge field of a particular spin, which, upon quantization,
gives rise to the Hilbert space of massless states of a given type.
This leads to the nontrivial matching condition on $h$
called in \cite{KV} admissibility condition.

As an illustration let us recall the standard argument
used to classify pure  supergravity models.
The first step is to say that supergravity results from gauging the
supersymmetry algebra with the generators
of (Minkowski or $AdS$) translations $P^a$, Lorentz transformations
$L^{ab}=-L^{ba}$, supertransformations $Q^i_\mu$ ($i=1\ldots \N$,
index $\mu$ is spinorial)
and global symmetries $T^{ij}=-T^{ji}$. One concludes that
any SUGRA model has to describe a set of gauge fields which contains
spin 2 massless field (graviton) described by the frame 1-form
$h^a$ and Lorentz connection 1-form $\go^{ab}=-\go^{ba}$
identified with the gauge fields corresponding to
$P^a$ and $L^{ab}$, spin 3/2 massless fields (gravitino)
described by 1-form spinors $\psi^i_\mu$ which are gauge fields for
$Q^i_\mu$, and spin 1 massless gauge fields $A^{ij}$ which correspond
to $T^{ij}$.

The second step is to check
whether the supersymmetry algebra  admits a
unitary  representation with exactly these sets of massless
states plus, may be, some lower spin states which are not described by
gauge fields. The answer is well known (see e.g. \cite{PVN}).
For $4d$ SUGRA, for example, the result is that for $\N\leq 8$ there is a
massless supermultiplet with the highest spin 2 and the required pattern
of spins 3/2 and 1. (In fact, with the exception of $\N=7$ when
the corresponding supermultiplet turns out to be a $\N=8$ supermultiplet).
For $\N>8$ (i.e., with more than total 32 supercharges) one finds that
there is always a state of spin $s>2$ in every massless supermultiplet.
Since any such a field is a gauge field, this does not match the list of
gauge fields of the usual SUSY algebra.

As long as it was not known
that nontrivial theories which contain massless HS fields require a
curved background with nonzero cosmological constant in a most symmetric
vacuum \cite{FV1}, consistent theories of HS massless fields were
not believed to
exist and the appearance of HS fields in SUSY supermultiplets used
to ``rule out" supergravities with $\N>8$. A more constructive alternative
was to study if there
exist some extensions of the usual SUSY algebras such that a set of
corresponding gauge fields would match some of their massless unitary
representations. The analysis along these lines opens a way towards
infinite dimensional HS algebras and nontrivial HS gauge theories.
It was originally applied in \cite{KV} to the $d=4$ case.
In \cite{KV1} a full list of admissible $4d$ HS algebras was obtained.
The corresponding nonlinear HS theories were constructed at the level
of classical field equations in \cite{more}. The aim of this paper is
to extend this analysis to any $d$.

Let us now discuss the general case.
Let an action $S(q)$ depend on some variables
$q^\Omega (x) $ and  be invariant under  gauge transformations
\be
\label{gentr}
\delta q^\Omega = r^\Omega (q; \gvep )\,,\qquad
\delta S =0\,,
\ee
where $\gvep^i (x)$ are infinitesimal gauge parameters. The
gauge transformation $r^\Omega (q; \gvep )$  contains the
variables $q^\Omega (x)$ and parameters $\gvep^i (x)$ along with
their derivatives.

Let $q^\Omega_0$ be some solution of the field equations
\be
\label{feq}
\f{\delta S}{\delta q^\Omega }\Big |_{q=q_0} = 0\,.
\ee
Perturbative analysis assumes that
$q^\Omega $ fluctuates near $q^\Omega_0$, i.e.
\be
q^\Omega = q^\Omega_0 + \eta q^\Omega_1\,,
\ee
where $\eta$ is some small expansion parameter.
A vacuum solution $q^\Omega_0$ is not invariant
under the gauge transformation (\ref{gentr}). One can however
address the question if there are some nonzero symmetry
parameters $\gvep^i_{gl} (x)$ which  leave the vacuum solution
invariant. Usually, if such a leftover symmetry exists at all,
the $x$-dependence of  $\gvep^i_{gl} (x)$ turns out to be fixed
in terms of values of $\gvep^i_{gl}(x_0)$ at any given point
$x_0$. This is why the leftover symmetries  are
global symmetries. The vacuum is called symmetric if a number of
its global symmetries is not smaller than
the number of local symmetries in the model, i.e.
$\gvep^i_{gl}(x_0)$ for some fixed $x_0$ are modules of a
global symmetry of the vacuum.

By definition of the global symmetry
transformation, $\delta_{gl} q^\Omega_0
=r^\Omega (q_0 , \gvep_0 )=0$. As a result
\be
\label{gltr}
\delta^{gl} q_1^\Omega =\delta^{gl}_{ 0} q_1^\Omega
+ o(\eta )\,,\qquad
\delta^{gl}_{0} q_1^\Omega=\f{\delta
r^\Omega (q,\gvep_0,\ldots)}{\delta q^\Lambda } \Big |_{q=q_0}
q_1^\Lambda\,,
\ee
where $o(\eta)$ denotes terms of higher orders in $q_1$.
The $\eta$-independent part of the transformation
(\ref{gltr}) acts linearly on the fluctuations $q^\Omega_1$ in a way
independent of a particular form of the nonlinear part of the
transformations (which is
important for the analysis of interactions, however).
Starting with some closed set of nonlinear gauge
transformations\footnote{These algebras may  form so called
open algebras with the commutators containing additional
``trivial'' transformations proportional to the
left hand sides of the field equations.}, the global symmetry
forms some (may be open) algebra as well. Expanding the full action as
\be
S= S_0 (q_0) + \eta^2 S_2 (q_0, q_1) + o(\eta^2)
\ee
(as usual, the term linear in $\eta$ is absent because
$q_0$ solves the field equations (\ref{feq}))
one observes that  the global symmetry $\delta^{gl}_0 q_1^\Omega$
is a symmetry of the free action   $S_2$ bilinear in the fields $q_1$.
Once the full theory is consistent, single-particle
quantum states of the theory span a Hilbert space, which
 forms a bounded energy unitary
$h$-module. It should decompose into a sum of unitary $g$-modules
of the space-time symmetry $g\subset h$. The decomposition into
UIRREPs of $g$ gives a list of particles described by the free
action $S_2$ as well as their quantum numbers like spins and
masses.

A particularly useful way to describe models with local symmetries
is by using $p$-form gauge fields \be \go^i (x) = dx^{\un_1}\wedge
\ldots\wedge dx^{\un_p} \go_{\un_1\ldots \un_p}^i (x)\,. \ee In the
recent paper \cite{ASV} it was shown how this approach can be used
to describe any $AdS_d$ massless representation associated with a
gauge field. The rule is very simple. A massless module $\H (E_0
(\bs) , \bs)$ with the $o(d-1)$ weight $\bs$ corresponding to a
finite dimensional $o(d-1)$-module formed by $o(d-1)$ traceless
tensors with the symmetry properties of a Young tableau

\sbox{\toch}{\circle*{1}}
\sbox{\gor}{\line(1,0){5}}
\sbox{\ver}{\line(0,1){5}}
\bee   
\label{vacuum}
\begin{picture}(150,125)(0,5)%
{
\put(0,0){\line(0,1){40}}
%
\multiput(3,0 )(7,0){04}{\usebox{\toch}}%
\multiput(3,09)(7,0){06}{\usebox{\toch}}%
\multiput(3,18)(7,0){09}{\usebox{\toch}}%
\multiput(0,30)(5,0){13}{\usebox{\gor}}%
\multiput(0,30)(5,0){14}{\usebox{\ver}}%
\multiput(0,35)(5,0){13}{\usebox{\gor}}%
\multiput(0,35)(5,0){14}{\usebox{\ver}}%
\multiput(0,40)(5,0){13}{\usebox{\gor}}%
\multiput(0,40)(5,0){14}{\usebox{\ver}}%
\multiput(0,45)(5,0){13}{\usebox{\gor}}%
\multiput(0,45)(5,0){14}{\usebox{\ver}}%
\multiput(0,50)(5,0){13}{\usebox{\gor}}%
\multiput(0,50)(5,0){14}{\usebox{\ver}}%
\multiput(0,55)(5,0){13}{\usebox{\gor}}%
\multiput(0,55)(5,0){14}{\usebox{\ver}}%
\multiput(0,60)(5,0){13}{\usebox{\gor}}%
\multiput(0,60)(5,0){14}{\usebox{\ver}}%
\put(28,21){${\tilde  s}_{2}$}
\put(75,43){${\tilde p}_2$}
\multiput(0,65)(5,0){19}{\usebox{\gor}}%
\multiput(0,65)(5,0){20}{\usebox{\ver}}%
\multiput(0,70)(5,0){19}{\usebox{\gor}}%
\multiput(0,70)(5,0){20}{\usebox{\ver}}%
\multiput(0,75)(5,0){19}{\usebox{\gor}}%
\multiput(0,75)(5,0){20}{\usebox{\ver}}%
\multiput(0,80)(5,0){19}{\usebox{\gor}}%
\multiput(0,80)(5,0){20}{\usebox{\ver}}%
\multiput(0,85)(5,0){19}{\usebox{\gor}}%
\multiput(0,85)(5,0){20}{\usebox{\ver}}%
\multiput(0,90)(5,0){19}{\usebox{\gor}}%
\multiput(0,90)(5,0){20}{\usebox{\ver}}%
\multiput(0,95)(5,0){19}{\usebox{\gor}}%
\multiput(0,95)(5,0){20}{\usebox{\ver}}%
\multiput(0,100)(5,0){19}{\usebox{\gor}}%
\multiput(0,100)(5,0){20}{\usebox{\ver}}%
\multiput(0,105)(5,0){19}{\usebox{\gor}}%
\multiput(0,105)(5,0){20}{\usebox{\ver}}%
\multiput(0,110)(5,0){19}{\usebox{\gor}}%
\multiput(0,110)(5,0){20}{\usebox{\ver}}%
\multiput(0,115)(5,0){19}{\usebox{\gor}}%
\multiput(0,115)(5,0){20}{\usebox{\ver}}%
\multiput(0,120)(5,0){19}{\usebox{\gor}}%
\put(50,123){$ s$}
\put(105,90){$p$}
}
\end{picture}
\eee
can be described in terms of a $p$-form gauge field that
takes values in the finite dimensional $o(d-1,2)$-module
depicted by the $o(d-1,2)$ traceless Young tableau
\bee   
\label{blockd}
\begin{picture}(150,130)(0,5)%
{
\put(0,0){\line(0,1){40}}
\multiput(3,0 )(7,0){04}{\usebox{\toch}}%
\multiput(3,09)(7,0){06}{\usebox{\toch}}%
\multiput(3,18)(7,0){09}{\usebox{\toch}}%
\multiput(0,30)(5,0){13}{\usebox{\gor}}%
\multiput(0,30)(5,0){14}{\usebox{\ver}}%
\multiput(0,35)(5,0){13}{\usebox{\gor}}%
\multiput(0,35)(5,0){14}{\usebox{\ver}}%
\multiput(0,40)(5,0){13}{\usebox{\gor}}%
\multiput(0,40)(5,0){14}{\usebox{\ver}}%
\multiput(0,45)(5,0){13}{\usebox{\gor}}%
\multiput(0,45)(5,0){14}{\usebox{\ver}}%
\multiput(0,50)(5,0){13}{\usebox{\gor}}%
\multiput(0,50)(5,0){14}{\usebox{\ver}}%
\multiput(0,55)(5,0){13}{\usebox{\gor}}%
\multiput(0,55)(5,0){14}{\usebox{\ver}}%
\multiput(0,60)(5,0){13}{\usebox{\gor}}%
\multiput(0,60)(5,0){14}{\usebox{\ver}}%
\put(28,21){${\tilde  s_{2}}$}
\put(75,43){${\tilde p}_2$}
\multiput(0,65)(5,0){18}{\usebox{\gor}}%
\multiput(0,65)(5,0){19}{\usebox{\ver}}%
\multiput(0,70)(5,0){18}{\usebox{\gor}}%
\multiput(0,70)(5,0){19}{\usebox{\ver}}%
\multiput(0,75)(5,0){18}{\usebox{\gor}}%
\multiput(0,75)(5,0){19}{\usebox{\ver}}%
\multiput(0,80)(5,0){18}{\usebox{\gor}}%
\multiput(0,80)(5,0){19}{\usebox{\ver}}%
\multiput(0,85)(5,0){18}{\usebox{\gor}}%
\multiput(0,85)(5,0){19}{\usebox{\ver}}%
\multiput(0,90)(5,0){18}{\usebox{\gor}}%
\multiput(0,90)(5,0){19}{\usebox{\ver}}%
\multiput(0,95)(5,0){18}{\usebox{\gor}}%
\multiput(0,95)(5,0){19}{\usebox{\ver}}%
\multiput(0,100)(5,0){18}{\usebox{\gor}}%
\multiput(0,100)(5,0){19}{\usebox{\ver}}%
\multiput(0,105)(5,0){18}{\usebox{\gor}}%
\multiput(0,105)(5,0){19}{\usebox{\ver}}%
\multiput(0,110)(5,0){18}{\usebox{\gor}}%
\multiput(0,110)(5,0){19}{\usebox{\ver}}%
\multiput(0,115)(5,0){18}{\usebox{\gor}}%
\multiput(0,115)(5,0){19}{\usebox{\ver}}%
\multiput(0,120)(5,0){18}{\usebox{\gor}}%
\multiput(0,120)(5,0){19}{\usebox{\ver}}%
\multiput(0,125)(5,0){18}{\usebox{\gor}}%
\put(40,128){$ s-1$}
\put(100,90){$p+1$}
}
\end{picture}
\eee
which is obtained from (\ref{vacuum}) by cutting the right
(i.e., shortest) column
and adding the longest row. For example, to describe the
spin two massless gauge field which corresponds to the
$o(d-1)$ Young tableau
$\begin{picture}(7,15)
{
\multiput(0,0)(5,0){2}{\usebox{\gor}}%
\multiput(0,0)(5,0){3}{\usebox{\ver}}%
\multiput(0,5)(5,0){2}{\usebox{\gor}}%
}
\end{picture}
$$\,\,$,
one introduces 1-form gauge field taking values in the representation
$\begin{picture}(11,7)
{
\multiput(0,0)(0,5){2}{\usebox{\ver}}%
\multiput(0,0)(0,5){3}{\usebox{\gor}}%
\multiput(5,0)(0,5){2}{\usebox{\ver}}%
}
\end{picture}
$
      of $o(d-1,2)$ to be interpreted as
the gauge connection of $o(d-1,2)$. Its decomposition into
representations of the Lorentz algebra $o(d-1,1)$ gives rise to
the frame 1-form and Lorentz connection.

The sector of 1-forms is particularly important because the 1-form
gauge fields take values in some
(super)algebra $h$ with structure coefficients $f^i_{jk}$. The
field strength is
\be
R^i= d\go^i (x) + f^i_{jk} \go^j (x)\wedge \go^k (x)
\,,\qquad d=dx^\un \f{\p}{\p x^\un}\,.
\ee
As just mentioned, gravitation is described in
such a formalism by the components of the connection 1-form
$\go_0$ of the $AdS_d$ algebra $o(d-1,2)$. For background geometry
with nondegenerate metric, $\go_0(x)$ must be
nonzero because it contains a nondegenerate frame field as the
component of the gauge field $\go^i (x)$ associated with the
generator of space-time translations in $g$ (note that there
exists invariant way to impose this condition with the help of the
so called compensator formalism  \cite{SW,PV,5d}). Gauge invariant
Lagrangians for gauge fields of any symmetry type can be built in
terms of the field strengths of $\go$ analogously to the
MacDowell-Mansouri formulation for gravity \cite{MM}.

Such a formalism has several nice properties.
In particular, it allows a natural zero-curvature
vacuum solution $\go_0$ of the dynamical field equations such that
\be
\label{R0}
R^i (\go_0 ) =0\,.
\ee
For the case of gravity with the connection $\go_0$ taking values
in $o(d-1,2)$, this is just the equation for the $AdS_d$ space-time.

Any  vacuum solution (\ref{R0}) has $h$ as a global symmetry (we
disregard here possible global topological obstructions in less
symmetric locally isomorphic spaces,  extending, if necessary, the
problem to the universal covering space-time). Indeed, the equation
(\ref{R0}) is invariant under the gauge transformations \be
\label{de0} \delta \go_0^i (x)= D_0\gvep^i (x)\equiv d\gvep^i (x)
+f^i_{jk} \go_0^j (x)\wedge \gvep^k (x)\,. \ee To have a fixed
vacuum solution $\go^i_0 (x)$ invariant one requires \be \label{de}
\delta \go_0 =D_0 (\gvep^i(x) )\equiv d\gvep^i (x) +f^i_{jk} \go^j_0
(x)\wedge \gvep^k (x)=0\,. \ee The equation (\ref{de}) is formally
consistent because $D_0^2 =0$. As a result it determines all
derivatives of the 0-form $\gvep^i(x)$ in terms of its values
$\gvep^i(x_0)$ at any given point $x_0$ of space-time. The
corresponding parameters  $\gvep^i_0 (x)$, which solve (\ref{de}),
are fixed in terms of $\gvep^i_0 (x_0) \in h$. They describe global
symmetry $h$ of the vacuum (\ref{R0}). Let us note that the same
argument is true for any gauge transformations which differ from
(\ref{de0}) by terms proportional to the field strengths and/or
matter (non-gauge) fields which are all assumed to have zero VEVs.
Also, let us note that $p$-gauge forms have $(p-1)$-form gauge
parameters. However, as a consequence of the Poincare lemma only
1-form gauge fields give rise to nontrivial global symmetries with
0-form parameters.

Given Lie superalgebra $h$ we can check using the results of
\cite{ASV}  whether or not it is appropriate to describe some set
of HS gauge fields and, if yes, to find out the spectrum of spins
of this set. Note that the condition that the gauge fields of $h$
correspond to some set of massless fields is itself nontrivial,
imposing  rigid restrictions on  $h$. In particular, according to
\cite{ASV}, a result of the decomposition of  $h$ into irreducible
submodules of the space-time symmetry $o(d-1,2)$ has to contain
only finite dimensional\footnote{This guarantees that the gauge
fields corresponding to every $g$-submodule in $h$ have a finite
number of components. Note that usually HS models describe
infinite set of fields, each having a finite number of components  to
describe one or another particle.} representations of $o(d-1,2)$
of special types, namely those depicted by traceless
Young tableaux of   $o(d-1,2)$ that have two first rows of equal
length.

Suppose now that there is a consistent nonlinear
theory of massless gauge fields formulated in terms
of connections of some algebra $h$ plus, may be, some number
of fields described by 0-forms as well as by higher forms.
Consistent interactions
may deform a form of the transformation law (\ref{de0}) by some
terms proportional to the curvatures $R$ and/or matter fields.
In the vacuum with zero curvature and matter fields the deformation terms
do not contribute to the global symmetry transformation law.
As a result, if a consistent nonlinear theory exists, $h$ is the
global symmetry
algebra of such its vacuum solution. This  automatically
implies that the space-time symmetry $g\subset h$ is also the global
symmetry of the model. It is one of the advantages of the formulation
in terms of gauge connections  that it makes global
symmetries of the model manifest including the usual space-time
symmetries. Note that this is achieved in a coordinate independent
way because it is not necessary to know a particular form of the vacuum
gauge connection $\go_0$. Instead, it is enough to impose (\ref{R0})
along with the condition that the frame field is nondegenerate
(see \cite{5d} for more details for the example of gravity).

Once $h$ is the global symmetry algebra of a hypothetical consistent
HS theory it has to obey the admissibility condition \cite{KV} that
there should be a unitary $h$--module which describes a list of
quantum single-particle states corresponding to all HS gauge fields
described in terms of the connections of $h$. If no such a module
exists, there is no chance to find a nontrivial consistent (in
particular, free of ghosts) theory that admits $h$ as a symmetry of
its most symmetric vacuum. On the other hand, once some
(super)algebra $h$ satisfying the admissibility condition is found,
the pattern of the appropriate unitary module also contains the
information on the matter fields and higher form HS gauge fields
analogous to the matter and higher form fields in extended
supergravity supermultiplets. As a result, one obtains a list of
($p=0$) matter and ($p>1$)-form fields to be introduced to make it
possible to build a consistent theory.

We see that {\it a priori} not every extension $h$ of $g$ is a HS
algebra. In particular, it is interesting to check whether the
algebras considered in \cite{cali} satisfy the formulated
criteria. The admissibility condition is the necessary condition
for $h$ to underly some consistent HS theory. In practice, all
examples of algebras $h$ known so far, which satisfy the
admissibility condition, turned out to be vacuum symmetries of some
consistent HS theories. Let us note  that the admissibility
condition applies to the symmetry algebras of on-mass-shell
single-particle states. It therefore does not provide any criterion
on the structure of possible further extensions of the HS algebras
with the additional fields being pure gauge or auxiliary, i.e.,
carrying no degrees of freedom. Such algebras may have physical
relevance as off-mass-shell algebras giving rise to some auxiliary
field variables which are zero by virtue of dynamical field
equations.

\
\section{Simplest bosonic higher spin algebras}
\label{Simplest bosonic higher spin algebras}

The HS algebra used in \cite{d} is defined as follows.
Consider bosonic oscillators $Y_i^A$ with $i=1,2$
\be
\label{inv}
(Y_i^A)^\dagger = Y_i^A \,\qquad
\ee
satisfying the commutation relations
\be
\label{defr}
[Y_k^A , Y_j^B ] =2i C_{kj}\eta^{AB}\,,
\ee
where $ C_{ij}= - C_{ji}\,,\quad C^{ij}C_{ik} = \delta_k^j $.
The bilinear forms $\eta^{AB}$ and $C^{ij}$ are used to raise and
lower indices in the usual manner:
$
A^A = \eta^{AB} A_B
$,
$
a^i =C^{ij}a_j
$,
$
a_i =a^j C_{ji}\,.
$

The generators of $o(M,2)$  are
\be
\label{T}
T^{AB} = -T^{BA}=\f{1}{4i} \{ Y^{jA}\,, Y^B_j\}\,.
\ee
Also we introduce the generators of $sp(2)$
\be
\label{osp}
t_{kl} = \f{1}{4i} \eta_{AB} \{Y^A_k\,,Y^B_l \} \,.
\ee
The generators $T^{AB}$ and $t_{ij}$ commute to the oscillators
$Y^A_j$ as follows
\be
[T^{AB} , Y^C_j ] = \eta^{BC} Y^A_j - \eta^{AC} Y^B_j\,,\qquad
[t_{jk}, Y^A_n ] = C_{kn} Y^A_j + C_{jn} Y^A_k \,,
\ee
i.e., $T^{AB}$ generate $o(M,2)$ rotations of indices $A,B,\ldots$,
while $t_{ij}$ generate $sp(2)$ transformations of indices
$i,j,\ldots$
Because indices in (\ref{T}) and (\ref{osp}) are contracted
with the $sp(2)$ and $o(M,2)$ invariant forms, respectively,
one finds that
\be
[T^{AB} ,t_{ij}] =0\,,
\ee
i.e., $o(M,2)$ and $sp(2)$ are Howe dual \cite{Howe}.
Note that the following identity is true
as a consequence of (\ref{T}) and (\ref{osp})
\be
\label{cascas}
C_2 \equiv - \half T^{AB}T_{AB} = -\f{1}{4} (M^2 -4)
-\f{1}{2} t^{ij} t_{ij}\,.
\ee

Consider the associative Weyl  algebra $A_{M+2}$ spanned by
the elements
\be
\label{exp}
f(Y) = \sum_{m}f_{A_1 \ldots A_m}^{i_1\ldots i_m}
Y_{i_1}^{A_1}\ldots Y_{i_m}^{A_m}\,,
\ee
i.e. $A_{M+2}$ is the enveloping algebra of the commutation
relations (\ref{defr}).
Consider the subalgebra
$S\subset A_{M+2}$ spanned by the $sp(2)$ singlets
\be
\label{ssp2}
f(Y)\in S\,:\qquad [t_{ij} , f(Y ) ] =0\,.
\ee
These conditions admit a nontrivial solution with
$f(Y)$ being an arbitrary function of the $o(M,2)$
generators (\ref{T}). Let $\hc$ be the Lie
algebra resulting from $S$  with the commutator
$[a,b]$ as the Lie product.
(In this notation, $2\!:\![M,2]$ refers to the dual pair
$sp(2)\oplus o(M,2)$, $h$ abbreviates ``higher", $c$ abbreviates
``centralizer", and 1 refers to a number of Chan-Paton indices
as explained in section \ref{Higher Spin Superalgebra}.)

The algebras $S$ and $\hc$ contain two-sided ideals $I$
spanned by the elements of the form
\be
\label{I}
g=t_{ij}g^{ij}\,,\qquad g\in S
\ee
where $g^{ij} (Y)$ behaves as a  symmetric
tensor of $sp(2)$, i.e.,
\be
[t_{ij}\,,g^{kl}]=
\delta_{j}^{k} g_i{}^{l}+
\delta_{i}^{k} g_j{}^{l}+
\delta_{j}^{l} g_i{}^{k} +
\delta_{i}^{l} g_j{}^{k}
\ee
(note that $t_{ij}g^{ji}= g^{ji}t_{ij}$).
Actually, from (\ref{ssp2}) it follows that
$fg,\, gf \in I$\, $\forall f\in S$, $g\in I$.
{}From the definition (\ref{osp}) of $t_{ij}$
one concludes that the ideal $I$ takes away all traces
of the $o(M,2)$ tensors so that the algebra
$S/I$ has only traceless
$o(M,2)$ tensors in the expansion (\ref{exp}).

The $sp(2)$ invariance condition  (\ref{ssp2}) is equivalent to
\be
\Big(Y^{Ai} \f{\p}{Y^A_j}  + Y^{Aj} \f{\p}{Y^A_i}
\Big) f(Y) =0\,.
\ee
For the expansion (\ref{exp}), this condition implies
that the coefficients
$f_{A_1 \ldots A_m\,C_1 \ldots  C_n}^{
\overbrace{1\ldots 1}^m \overbrace{2\ldots 2}^n}$
are nonzero only if $n=m$
and that symmetrization over any $m+1$ indices among
${A_1, \ldots A_m\,,C_1, \ldots  C_m}$ gives zero.
This implies that the coefficients
$f_{A_1 \ldots A_m\,C_1 \ldots  C_m}^{
\overbrace{1\ldots 1}^m \overbrace{2\ldots 2}^m}$
have the symmetry properties of the two-row rectangular
Young tableau
\begin{picture}(45,25)
{
\put(00,00){\line(1,0){40}}%
\put(00,05){\line(1,0){40}}%
\put(00,10){\line(1,0){40}}%
\put(00,00){\line(0,1){10}}%
\put(05,00.0){\line(0,1){10}} \put(10,00.0){\line(0,1){10}}
\put(15,00.0){\line(0,1){10}} \put(20,00.0){\line(0,1){10}}
\put(25,00.0){\line(0,1){10}} \put(30,00.0){\line(0,1){10}}
\put(35,00.0){\line(0,1){10}} \put(40,0.0){\line(0,1){10}}
}
\put(12,10.9){\scriptsize  $m$}
\end{picture}
(for more details see Appendix). Thus, the algebras $S$ and $\hc$
decompose into direct sums of $o(M,2)$--modules described by various
two-row rectangular Young tableaux. The algebra $S/I$ is spanned by
the elements with traceless $o(M,2)$ tensor coefficients which have
symmetry properties of two-row Young tableaux. The Lie algebra $\hc
/I$ was identified by Eastwood with the conformal HS algebra in $M$
dimensions in \cite{East} where its realization in terms of the
enveloping algebra of $o(M,2)$ was used. In \cite{d} the algebra
$\hc /I$ was called $hu(1/sp(2)[M,2])$. To simplify notations we
call this algebra $\hu$ in this paper. More precisely, $\hu$ is the
real Lie algebra singled out by the reality condition \be f(Y)\in
\hu :\qquad f(Y)\in \hc /I\,,\quad f^\dagger (Y) = - f(Y)\,, \ee
where $\dagger$ is the involution (\ref{inv}) extended to a generic
element by the standard properties $(fg)^\dagger = g^\dagger
f^\dagger $, $(\lambda g)^\dagger = \bar{\lambda} f^\dagger $ for
$f,g \in \hc /I, \lambda \in {\bf C}$. Note that $\hu$ contains the
$o(M,2)$ generators $T^{AB}$ (\ref{T}).

Let us now consider another Lie algebra $\huo$
resulting from the analogous construction with the
additional Clifford elements $\phi^A $  satisfying the
anticommutation relations
\be
\label{cl1}
\{\phi^A , \phi^B \} = -2\eta^{AB}\,.
\ee
Now the generators of $o(M,2)$ are realized as
\be
\label{TS}
T^{AB}=-T^{BA}=\f{1}{4i}\{Y^{jA} , Y_j{}^B\}  -\f{1}{4}[\phi^A , \phi^B ]\,.
\ee
They commute to the (super)generators of $osp(1,2)$
\be
\label{osp1}
t_j = \f{1}{2} Y^A_j \phi_A\,,\qquad
t_{nm} = i \{t_n , t_m \} =\f{1}{4i} \eta_{AB}\{ Y^A_n , Y^B_m\}\,.
\ee

The associative Weyl-Clifford algebra
$A_{M+2,M+2}$ is spanned by the elements
\be
\label{genel}
f(Y,\phi) = \sum_{m,n,k}
f_{A_1 \ldots A_m\,,B_1\ldots B_n ;C_1\ldots C_k}
Y_{1}^{A_1}\ldots Y_{1}^{A_m}
Y_{2}^{B_1}\ldots Y_{2}^{B_n}
\phi^{C_1}\ldots \phi^{C_k}\,,
\ee
where the coefficients
$f_{A_1 \ldots A_m\,,B_1\ldots B_n ;C_1\ldots C_k}$ are totally
symmetric (separately) in the indices $A_1 \ldots A_m$ and
$B_1\ldots B_n$ and totally antisymmetric in the indices
$C_1\ldots C_k$. Its subalgebra $\S$ is spanned by the elements
which have zero graded commutator with the (super)generators
of $osp(1,2)$, i.e.
\be
\label{superc}
f\in \S\,:\quad f(Y,\phi) t_j =t_j  f(Y,-\phi)\,.
\ee
{}From (\ref{osp1}) it follows  that any $f\in \S$ is
 $sp(2)$ invariant. As a result,  $n=m$ and
$f_{A_1 \ldots A_m\,,B_1\ldots B_m ;C_1\ldots C_k}$ has the
symmetry properties of a two-row rectangular Young tableau
with respect to the indices $A_1 \ldots A_m$ and $B_1\ldots B_m$,
i.e. symmetrization over any $m+1$ indices $A_i$ and $B_j$
gives zero. The condition (\ref{superc})  implies
\be
(k+1)f_{A_1 \ldots A_m\,,B_1\ldots B_m ;A_{m+1} C_1\ldots C_k}
 =i
(m+1)f_{A_1 \ldots A_{m+1}\,,B_1\ldots B_m C_1; C_2\ldots C_k}\,,
\ee
where total (anti)symmetrization over the indices $(C)A$ is
understood. Its general solution is
\bee
\label{gens}
f_{A_1 \ldots A_m\,,B_1\ldots B_m ; C_1\ldots C_k}&=&
g_{A_1 \ldots A_m\,,B_1\ldots B_m | C_1\ldots C_k}\nn\\
&+& i \theta (k-2)\f{(m+1)^2}{m+k}
g_{A_1 \ldots A_m C_1\,,B_1\ldots B_m  C_2 | C_3\ldots C_k}\,,
\eee
where
$g_{A_1 \ldots A_m\,,B_1\ldots B_m | C_1\ldots C_k}$ is an arbitrary
tensor that has symmetry properties of the Young tableau
\bee
\begin{picture}(45,50)
{
\put(00,40){\line(1,0){40}}%
\put(00,35){\line(1,0){40}}%
\put(00,30){\line(1,0){40}}%
\put(00,25){\line(1,0){05}}%
\put(00,20){\line(1,0){05}}%
\put(00,15){\line(1,0){05}}%
\put(00,10){\line(1,0){05}}%
\put(00,05){\line(1,0){05}}%
\put(00,00){\line(1,0){05}}%
\put(00,00){\line(0,1){40}} \put(05,00.0){\line(0,1){40}}
\put(10,30.0){\line(0,1){10}}
\put(15,30.0){\line(0,1){10}} \put(20,30.0){\line(0,1){10}}
\put(25,30.0){\line(0,1){10}} \put(30,30.0){\line(0,1){10}}
\put(35,30.0){\line(0,1){10}} \put(40,30.){\line(0,1){10}}
}
\put(22,42.){\scriptsize  $m$}
\put(-30,20){\scriptsize  $k+2$}
\end{picture}.
\label{kryuk}
\eee
In other words $g_{A_1 \ldots A_m\,,B_1\ldots B_m | C_1\ldots C_k}$
 is totally symmetric in the indices $A_1 \ldots A_m$ and $B_1\ldots
 B_m$,
totally antisymmetric in the indices $C_1\ldots C_k$ and such that
symmetrization over any $m+1$ indices gives zero.

Let $\hco$ be the Lie algebra isomorphic to $\S$
as a linear space, with the commutator (not graded commutator!) in $\S$
as the Lie bracket. The factorization over the ideal $I$
spanned by the elements of $\hco$ which are themselves
proportional to the generators of $osp(1,2)$
implies that the tableaux (\ref{kryuk}) are traceless.
The resulting algebra $\huo$ decomposes as a linear
space into the direct sum of traceless representations of $o(M,2)$
which have the symmetry properties of the Young tableaux (\ref{kryuk}).
More precisely,  $\huo$ is the real  Lie algebra
spanned by the elements satisfying
\be
\label{reali}
(f(Y,\phi ))^\dagger = -f(Y, \phi)
\ee
at the condition that (\ref{inv}) is true along with
\be
\label{phidag}
(\phi_A)^\dagger = -\phi_A\,.
\ee
(Let us note that the involution $\dagger$ can be realized
in the usual manner as $(a)^\dagger = \phi^0 \phi^{M+1} a^+
\phi^{M+1}\phi^0$, where $a^+$ is some Hermitian conjugation with respect
to a positive-definite form. Note that a sign on the right hand side
of (\ref{phidag}) is chosen so that
the space-like components of $\phi_A$ could
be realized by hermitean matrices).

{}From the formula
\begin{equation}
\label{kind}
\gep_{{n_1}\ldots n_{M+2}}\epsilon_{{m_1}\ldots m_{M+2}}=
\sum_{p}(-1)^{\pi(p)}\eta_{n_1 m_{p (1)}}\ldots \eta_{n_{M+2} m_{p
({M+2})}}\,,
\end{equation}
where summation is over all permutations $p$ of indices
$m_i$ and ${\pi (p)}= 0$ or $1$ is the oddness of the permutation
$p$, it follows that any traceless tensor with the symmetry properties
of a Young tableau, which contains two columns with more that $M+2$
cells, is identically zero\footnote{By virtue of (\ref{kind})
one proves that a tensor twice dual to the
original one in the two groups of antisymmetrized indices must vanish
because at least one of the metric
tensors on the right hand side of (\ref{kind}) will be contracted
with a pair of indices of the dualized tensor. By double
dualization one gets back the original tensor which is therefore also zero.}.
{}From here it follows that only
Young tableaux with up to $M$ cells in the first column appear
among the $o(M,2)$ representations contained in $\huo$, having
at least two nonzero columns. It turns out however that
elements described by the one-column Young tableaux of height 1 and
$M+1$ belong to the ideal $I$
and, therefore, do not appear among the elements of
$\huo$.  This fact is the content of {\it Lemma 3.1}
of section \ref{Projection technics}.
Let us mention that the feature that factoring out elements
proportional to $t_i$ may imply some factorization beyond
only taking away traces is
because Clifford algebra is finite dimensional. In other words,
 when fermions are present, the factorization
over the ideal $I$ may  take away  traces along with
some traceless elements.

To summarize, as $o(M,2)$-module $\huo$
decomposes into the sum of all finite dimensional
$o(M,2)$-modules described by various traceless Young tableaux
(\ref{kryuk}) except for those with the first column
of heights $1$ or $M+1$. (The trivial tableau with
no cells and its dual described by the one-column tableau with
$M+2$ cells are included).
Each allowed irreducible  $o(M,2)$-module appears in one copy.

The element \be \label{Gamma} \Gamma =(i)^{\half (M-2)(M-3)}
\phi^0\phi^1\ldots \phi^{M+1} \ee satisfies \be \label{Gcom}
\Gamma \phi^A = (-1)^{M+1} \phi^A \Gamma \,,\qquad \Gamma^2=Id\,
\ee and \be \label{gdag} \Gamma^\dagger =  \Gamma \,. \ee As a
result, the projectors \be \label{pipm} \Pi_\pm = \half (1\pm
\Gamma )\, \ee are Hermitian \be \label{12} (\Pi_\pm )^\dagger
=\Pi_\pm\,. \ee According to (\ref{Gcom}), the elements $\Gamma$
and $\Pi_\pm$ are central for odd $M$. As a result, analogously to
the case of usual Clifford algebra, $\huo$
decomposes for odd $M$ into direct sum of two subalgebras singled out
by the projectors $\Pi_\pm$
\be
\label{oddec}
\huo=
\hue\oplus \hue\,.
\ee
Here $\hue$ is the subalgebra of $\huo$
spanned by the elements even in $\phi$, $f(Y,-\phi )$ = $f(Y,\phi
)$, described by various Young tableaux (\ref{kryuk}) with even
numbers of cells. Note that for the particular case of $M=3$,
which corresponds to $AdS_4$, the algebra $hu^E(1|(1,2)\!\!:\!\![3,2])$
as a $o(M,2)$-module contains only rectangular two-row Young
tableaux. As will be shown in section \ref{Spinorial realization},
in agreement with the $4d$ results of \cite{KV1}, this is the
manifestation of the isomorphism $hu^E(1|(1,2)\!\!:\!\![3,2])\sim
hu(1|2\!\!:\!\![3,2])$.

By definition of $\hue$, its elements commute with $\Gamma$ and
$\Pi_\pm$. For even $M$ one can therefore define two algebras
$\hupm$ spanned by the elements of the form \be \label{chir1} b\in
\hupm : \,\quad b= \Pi_\pm a\,,\qquad a\in \hue\,. \ee Note  that
$\hupm$ are not subalgebras of $\hue$ because their elements do not
satisfy (\ref{superc}) in the sector of the $osp(1,2)$ supercharges
$t_j$. Elements of $\hupm $ are even rank $o(M,2)$ tensors such that
the tensors, described by the Young tableaux with the heights $p$
and $M+2-p$ of the  first column,  are dual to each other. In
particular, the $o(M,2)$ generators in $\hupm $ are $\Pi_\pm T_{AB}$
where $T_{AB}$ are the generators (\ref{T}). For $M+2=4q$, the rank
$2q$ $o(M,2)$ tensors are (anti)selfdual (for $M=4q$, rank $2q+1$
tensors do not belong to $\hue$).

We will call the algebras $\hue$ and $\hupm $ ($M$ is even)
- type $A$ and type $B$ HS algebras, respectively.
To summarize, let us list the gauge fields associated with the
HS algebras defined in this section.

The gauge fields of $\hu$ are 1-forms
$\go_\un^{A_1\ldots A_n,B_1\ldots B_n}$
carrying representations of $o(M,2)$ described by
 various two-row traceless rectangular Young tableaux
of lengths $n=0,1,2\ldots$. As shown in \cite{5d,d}, these
describe totally symmetric massless fields in $AdS_{M+1}$,
i.e., the lowest energy subspace of the corresponding
UIRREP of $o(M,2)$
is described by the rank $n+1$ totally symmetric traceless
tensors of $o(M)$.

The gauge fields of $\huo$ are 1-forms
$\go_\un^{A_1\ldots A_n,B_1\ldots B_n,C_1\ldots C_m }$
carrying representations of $o(M,2)$ described by
 various  traceless Young
 tableaux having two rows of equal length and one column
of any height $m\leq M$. There are two degenerate cases
of 1-forms carrying totally antisymmetric representations
of zero or maximal ranks $\go_\un^{D_1\ldots D_n }$ with $n=0$ or $M+2$
(while the cases with $n=1$ or $n=M+1$ are excluded).
According to the results of \cite{ASV}, the gauge fields
$\go_\un^{A_1\ldots A_n,B_1\ldots B_n,C_1\ldots C_m }$
with $n\geq 1$  describe
massless fields in $AdS_{M+1}$ corresponding to the
UIRREPs of $o(M,2)$ with the lowest energy states which
form representations of $o(M)$ described by  the traceless Young
tableaux having one row of length $n+1$
and one column of height $\min( m+1, M-m-1)$
\bee
\label{lkr}
\begin{picture}(40,50)
{
\put(00,35){\line(1,0){40}}%
\put(00,30){\line(1,0){40}}%
\put(00,25){\line(1,0){05}}%
\put(00,20){\line(1,0){05}}%
\put(00,15){\line(1,0){05}}%
\put(00,10){\line(1,0){05}}%
\put(00,05){\line(1,0){05}}%
\put(00,00){\line(1,0){05}}%
\put(00,00){\line(0,1){35}} \put(05,00.0){\line(0,1){35}}
\put(10,30.0){\line(0,1){05}}
\put(15,30.0){\line(0,1){05}} \put(20,30.0){\line(0,1){05}}
\put(25,30.0){\line(0,1){05}} \put(30,30.0){\line(0,1){05}}
\put(35,30.0){\line(0,1){05}} \put(40,30.){\line(0,1){05}}
}
\put(12,37.){\scriptsize  $n+1$}
\put(-100,20){\scriptsize  $\min(m+1, M-m-1)$}
\end{picture}.
\eee
The degenerate cases of $o(M,2)$ singlet 1-forms
$\go_\un^{A_1\ldots A_n,B_1\ldots B_n,C_1\ldots C_m }$
with $n=m=0$ and $n=1$, $m=M$
correspond to two spin 1 fields.
The fields with $n=1, m=0$ and $n=1, m=M-2$ describe two
graviton-like spin 2 fields.

The gauge fields of $\hue$ are 1-forms
$\go_\un^{A_1\ldots A_n,B_1\ldots B_n,C_1\ldots C_m }$ with even $m$.
The corresponding lowest energy representations of $o(M)\subset
o(M,2)$ are described by the hook Young tableaux (\ref{lkr}) with an
odd number of cells in the first column.

The type $B$ chiral algebras $\hupm$ ($M$ is
even) give rise to gauge fields $\go_\un^{A_1\ldots A_n,B_1\ldots
B_n,C_1\ldots C_m }$ with $m=l$ and $m = M -l-2$ related by
the (anti)selfduality conditions. In particular, the fields with
$m= 2q $ for $M=4q+2$ are (anti)selfdual, i.e. the corresponding
lowest energy representations of $o(4q+2)$ are described by
(anti)selfdual Young tableaux with the first column of height
$2q+1 $. Note that type $B$ systems can be realized in
odd-dimensional space-times $AdS_{M+1}$ which include the cases of
$AdS_5$ and $AdS_{11}$ being of special interest from the
superstring theory perspective.

\section{Factorization by projection}
\label{Projection technics}

To simplify the construction of HS algebras it is
convenient to use the projection formalism analogous to that
used in \cite{5d,5ds} for the analysis of $5d$ HS models.
The idea is that it is easy to factor out terms proportional
to one or another set of operators $a_i$ if there is some
element $ \Delta $ such that $a_i \Delta = \Delta  a_i =0$.
Let $C$ be the centralizer of $a_i$, i.e., $f\in C :
[a_i , f]=0 $. Suppose that
$\Delta$ also commutes with all elements of $C$, which is usually
automatically true because $\Delta$ is in a certain sense built of
$a_i$. Then elements $f\Delta = \Delta f$, $f\in C$ span $C/I$ where
$I$ consists of such  $g\in C$ that
$g=g^l a_i$ or $g=a_i g^r$ for some $i$.

A little complication is that in many cases $\Delta$ does not
belong to the original algebra because $\Delta^2$ does not exist
(diverges). As a result, the space of elements $f\Delta = \Delta
f$ forms a module of the original algebra rather than an algebra
with respect to the original product law. However, one can
redefine the product law of the elements of $g=f\Delta $
appropriately provided that there is an element $G$ such that \be
\label{dgd} \Delta G \Delta = \Delta. \ee One simply defines \be
\label{circ} g_1\circ g_2 = g_1 G g_2\,. \ee The new product is
associative and reproduces the product law in $C/I$ so that \be
\label{prodred} g_1\circ g_2 =f_1 f_2 \Delta\,,\qquad g_i = f_i
\Delta\,. \ee Note that $G$ is not uniquely defined because, with
no effect on the final result, one can add to $G$ any
 terms $a_i p$ and $q a_i$ with $p$ and $q$
annihilating $\Delta w$. In fact, the role of $G$ is auxiliary
because  one can simply use $(\ref{prodred})$ as
a definition of the product law in $C/I$.

This situation can be illustrated by the example of the
algebra of differential operators. Consider
differential operators with polynomial coefficients
of one variable $x$.  Its generic element is
$a(x,x^\prime )=
\sum_{n,m=0}^\infty a_{n,m} x^n \delta^m (x-x^\prime )$ with a finite
number of nonzero coefficients $a_{n,m}$ and
\be
\delta^m (x-x^\prime ) = \f{\p^m}{(\p x)^m}\delta(x-x^\prime )\,.
\ee
We consider simultaneously the case of usual commuting variable
$x$ and the Grassmann case with
$x^2=0$. (Recall that in the latter case
$\delta(x )= x$.) The product law is
defined as usual by $(ab)(x_1,x_2 ) = \int dx_3 a(x_1,x_3 ) b(x_3 , x_2 )$.
Consider the subalgebra
$C$ spanned by the elements which commute with
$x$ (i.e. $x\delta (x-x^\prime ))$. It is spanned by polynomials
of $x$, i.e.  $a\in C : a=\sum_{n=0}^\infty a_{n} x^n
\delta (x-x^\prime )$.
Obviously, the ideal $I$ is
spanned by the elements proportional to $x$ which are various elements
$a$ with zero constant term. The quotient algebra $C/I$ is
spanned by constants $ a_0 \delta (x-x^\prime )$.

Let us now do the same with the aid of projector. $\Delta$
is obviously the delta function
\be
\Delta (x, x^\prime ) = \delta(x) \delta (x^\prime )\,.
\ee
It satisfies $x\Delta = \Delta x = 0$. But $\Delta^2=\Delta \delta(0)$
with $\delta (0)=0$ in the Grassmann case and $\delta (0)=\infty$ in the
commuting case. Note that this is not occasional but
is a consequence of the original algebra properties. Actually, suppose
that $\Delta^2$ is well-defined. Then $\Delta^2$ would satisfy the
same properties $x\Delta^2 = \Delta^2 x = 0$ and one could expect that
$\Delta^2 = \alpha\Delta$ with some coefficient $\alpha$. If $\ga \neq 0$
or $\infty$, upon appropriate rescaling, $\Delta$ could
be defined as a projector. However, this cannot be true because, formally,
$\Delta^2$ = $\Delta [\f{\p}{\p x} , x ] \Delta =0$. Therefore, either
$\ga=0$ (Grassmann case) or $\Delta^2 $ makes no sense
(commuting case of $\ga = \infty$). To redefine the product law according
to (\ref{dgd}), (\ref{circ}) one can set
\be
G(x,x^\prime )=1\,.
\ee
There is an ambiguity in the choice of $G$. Any
$G^\prime (x,x^\prime )= G(x,x^\prime )
 +\sum_{n,m\geq 0} a_{n,m} x^n x^{\prime m}$ with $a_{0,0}=0$
is equally good. Note that neither $\Delta$ nor $G$ belong to the
original algebra of differential operators.

In the case of $\hc$ the operators $a_i$ identify
with the generators $t_{ij}$ of $sp(2)$. Let us define the algebra
of oscillators $Y^A_i$ as the star product algebra with the
product law
\be
\label{star} f(Y)*g(Y) = (2\pi)^{-2(M+2)} \int
d^{2(M+2)} S d^{2(M+2)} T\,f(Y+S)g(Y+T)\exp i S^A_j T_A^j\,,
\ee
which is the integral formula for the associative Weyl product
(sometimes called Moyal product) of totally symmetrized products
of oscillators. Here $Y^A_i$, $S^A_j$ and $ T^A_j $ are usual
commuting variables while the non-commutativity of the oscillator
algebra results from the non-commutativity of the star product.
{}From (\ref{star}) the following relations follow \be
\label{basr} Y^A_j * g(Y) = (Y^A_j +i\f{\p}{\p Y_A^j} ) g(Y)
\,,\qquad
 g(Y) *Y^A_j = (Y^A_j -i\f{\p}{\p Y_A^j} ) g(Y)\,.
\ee

To apply the projection method we need an element $\Delta $
that satisfies
\be
t_{ij} *\Delta = \Delta *t_{ij} =0\,.
\ee
An appropriate ansatz is
\be
\label{z}
\Delta = \Phi ( z)\,,\qquad z= \f{1}{4} Y^A_i Y_{Aj} Y^{Bi}Y_B^j \,,
\ee
where $z$ is both $sp(2)$ and $o(M,2)$ invariant and, therefore,
\be
[\Delta , t_{ij} ]_* =0\,,\qquad
[\Delta , T_{AB} ]_* =0\,,
\ee
where we use notations $[a,b]_* =a*b -b*a$, $\{a , b\}_* = a*b +b*a$.
{}From (\ref{star}) one finds that the condition
$\{t_{ij} , \Delta \}_* =0$ gives
\be
\left ( Y^A_i Y_{Aj} - \f{\p^2}{\p Y^{Ai}\p Y_A^j} \right )\Phi =0\,.
\ee
It is elementary to check that it amounts to the differential equation
\be
\label{eq}
2z \Phi^{\prime\prime} +(M+1)\Phi^\prime -\Phi =0\,,
\ee
where $\Phi^\prime (z) = \f{\p}{\p z} \Phi (z)$.
For $M>0$ this equation admits
the unique solution analytic in $z$ which can be written
in the form
\be
\label{Delta}
\Phi (z) = \int_{-1}^1 ds (1-s^2 )^{\half(M-2)} \exp{s\sqrt{z}}\,.
\ee
The analyticity in $z$ is because the integration region
is compact and the measure is even under $s\to -s$ so that only
even powers of $\sqrt{z}$ contribute. It is elementary to check
that it satisfies (\ref{eq}) by partial integration over $s$.

Let $f \in \hc$, i.e. $[f,t_{ij} ]_*=0$.
According to the general scheme, $\hu$ is spanned by
 the elements
\be
f*\Delta =\Delta *f\,.
\ee
Note that $[\Delta , f]_* =0$ because $\Delta $ is some function
of $z$ (\ref{z}) which itself is a function of $t_{ij}$\footnote{It
is useful to observe that  any order $p$ polynomial of $z$ can be represented
as some other order $p$ star polynomial of $t_{ij} * t^{ij}$. As a result,
any polynomial of $z$ has zero star commutator with any $f\in
 \hc$.}. It does not belong to the algebra
since $\Delta*\Delta $ diverges. The product law in $\hu$
is defined by the general formula (\ref{prodred}).

The case of the algebra $\huo$ with Clifford generating elements
$\phi^A$ can be considered analogously. First one replaces the Weyl
star product (\ref{star}) with the Weyl-Clifford one \bee
\label{sstar} f(Y,\phi)*g(Y,\phi) = (2\pi)^{-2(M+2)}&{}&\ls\!\! \int
d^{2(M+2)} S d^{2(M+2)} Td^{M+2}\chi  d^{M+2}\psi
\exp( i S^A_j T_A^j -\chi^A\psi_A)\nn\\
&\times& f(Y+S,\phi+\chi)g(Y+T, \phi+\psi)\,
\eee
such that the relations
\be
\label{basrc}
\phi^A * X= \Big(\phi^A -\f{\p}{\p \phi_A}\Big )X\,,\qquad
X*\phi^A =X \Big(\phi^A -\f{\overleftarrow{\p}}{\p \phi_A}\Big )
\ee
are true in addition to (\ref{basr}).
Then one observes that the operator $\Delta_1$
\be
\Delta_1 = \Phi (z_1) \,,\qquad z_1 = \f{1}{4} Y^A_i Y_{Aj} Y^{Bi}Y_B^j
+\f{i}{2}Y^A_i \phi_A Y^{Bj} \phi_B\,
\ee
possesses the desired properties
\be
\label{ct}
t_i * \Delta_1 =\Delta_1 *t_i = 0\,,\qquad
\ee
\be
\label{td1}
t_{ij}*\Delta_1 = \Delta_1  *t_{ij} =0\,.
\ee
The form of $\Delta_1$ can be guessed as follows.
Since $\Delta_1 $ is
a Casimir operator of $osp(1,2)$ it is natural to expect that
it is some function of its quadratic Casimir operator
$z_1$ (note that the star commutators
with $t_i$ and $t_{ij}$ have a form of some first-order differential
operators so that their annulators form a ring).
On the other hand, since $t_{ij}$ is
$\phi$-independent, the $\phi$-independent part of the
condition (\ref{td1}) implies that the functional dependence of
$\Delta_1$ on $z_1$ must be the same as that of $\Delta (z)$ on $z$.

According to the general scheme, elements of $\hu$
can be represented as
\be
\Delta_1* a = a*\Delta_1
\ee
with various $a$ such that $[a, t_{i}]_*=0 $
(and, therefore,  $[a, \Delta_1 ]_* =0$). Now the factorization of
elements $a=b^{i}*t_{i}$ or $a=t_{i}*b^{i}$ is automatic.

In applications it is convenient to use a slightly different
realization of $\Delta_1$.
Let us introduce the operator
\be
\label{ldef}
L= -\f{1}{4}(M-2) -\f{i}{8}
\phi^A \phi^B Y_{Aj}Y_B^j= 1  -\f{i}{2} t_j*t^j\,.
\ee
It is easy to check that it has the following properties
\be
L*t_i = \f{1}{2}t_{ij}* t^j\,,\qquad t_i*L = -\f{1}{2}t^j *t_{ij}\,,
\ee
\be
\label{lpr}
L^2 =L-\f{1}{8} t_{ij}* t^{ij}\,,
\ee
\be
\label{tl}
[t_{ij} , L]_* =0\,.
\ee

{}From (\ref{lpr}) it follows that $L$ is a projector modulo
terms proportional to $t_{ij}$. In particular,
\be
L^2 * \Delta = L*\Delta \,\qquad
\Delta *L^2 = \Delta *L\,.
\ee
{}From (\ref{tl}) it follows that
\be
[L, \Delta]_* =0\,.
\ee
One observes that $L*\Delta $ has the same properties as $\Delta_1$.
Indeed, using that $\Delta $ is annihilated by $t_{ij}$ from the left and
from the right, one finds that $L*\Delta $ satisfies
\be
\label{iden}
t_i*(L*\Delta )= (\Delta *L) * t_i =0\,,\qquad
t_{ij}*(L*\Delta )= (L*\Delta )*t_{ij} =0\,.
\ee
One therefore can  represent  $\Delta_1$ as $L*\Delta $. The precise
relationship is
\be
\label{1212}
\Delta_1 = -\f{4}{M-2} L*\Delta
\ee
as one can conclude from the first line in (\ref{ldef})
by comparing the $\phi$ independent parts in $L*\Delta$ and
$\Delta_1$. Let us note that the formula (\ref{1212}) does not
work for $M=2$. This means that the operators $\Delta_1$
and $L*\Delta$  are essentially different for this case. Presumably
this is related to the fact that the algebra $o(2,2)$
is not simple, having two independent quadratic Casimir operators.

Since $L$ is a well-defined polynomial operator in the star
product algebra, this allows us to give a useful
alternative definition of the algebra
$\huo$ as follows. Let $a(Y,\phi)$ be an arbitrary
$sp(2)$ invariant element of the star product algebra
\be
\label{at}
[a,t_{ij}]_*=0\,.
\ee
Then $\huo$ is spanned by  the elements
\be
\label{LDAL}
x=L*\Delta *a*L\,.
\ee
First note that because of (\ref{at}) the same element $x$ can be
equivalently written in any of the following forms
\be
x=L*a*\Delta*L = L*a*L*\Delta = \Delta*L*a*L\,.
\ee
Using the identities (\ref{iden}) one proves that
$t_i *x= x*t_i=0$. Clearly, any two $a$ are equivalent if
they differ by terms  $t_i *b^i$ or $b^i *t_i$ with some $b^i$.

Let us now prove that the rank 1 and rank $M+1$
antisymmetric representations factor out of $\huo$

\vskip0.3cm
\noindent
{\it Lemma 3.1}
\be
\label{3.1}
\Delta* L*\phi^A *L= 0\,,\qquad
\Delta* L*\phi^{A_1}\ldots\phi^{A_{M+1}}  *L= 0\,.
\ee
\noindent
The proof is elementary. One writes
\be
\Delta* L*\phi^A *L= \Delta* L* \phi^A -
\f{i}{2}\Delta* L* \phi^A *t_j *t^j\,
\ee
and then observes that
\be
i\Delta* L* \phi^A *t_j*t^j =
i\Delta* L* \{\phi^A ,t_j\}_* *t^j =
-i \Delta* L* Y_j^A *t^j =
-i \Delta* L* [ Y_j^A  , t^j ]_* = 2 \Delta* L* \phi^A \,.
\ee
The second identity in (\ref{3.1}) follows from the first one
along with $\Gamma*L = L*\Gamma$.

Now  we are in a position to define simplest HS superalgebras
in any dimension.

\section{Higher spin superalgebras}
\label{Higher Spin Superalgebra}

To define a superalgebra which unifies $\hu$ and $\huo$ we add two
sets of elements $\chi_\mu $ and $\bar{\chi}^\mu$ which form
conjugated  spinor representations of the $o(M,2)$ Clifford algebra
($\mu , \nu\ldots$ are spinor indices), \be \label{clre} \chi_\mu *
\phi^A= \gga^A{}_\mu{}^\nu \chi_\nu \,,\qquad \phi^A* \bar{\chi}^\mu
=  \bar{\chi}^\nu\gga^A{}_\nu{}^\mu \,, \ee where $o(M,2)$ gamma
matrices  $\gga^A{}_\nu{}^\mu$ satisfy \be \gga^A{}_\nu{}^\rho
\gga^B{}_\rho{}^\mu + \gga^B{}_\nu{}^\rho \gga^A{}_\rho{}^\mu =
-2\delta_\nu^\mu \eta^{AB}\,, \ee and commute with $Y^A_i$ \be
\chi_\mu  *Y_i^A = Y^A_i * \chi_\mu  \,,\qquad \bar{\chi}^\mu
* Y^A_i = Y^A_i *  \bar{\chi}^\mu\,. \ee Introduce two
projectors $\Pi_1$ and $\Pi_2$ \be \Pi_1*\Pi_1 = \Pi_1\,,\qquad
\Pi_2*\Pi_2 = \Pi_2\,,\qquad \Pi_1 *\Pi_2 = \Pi_2*\Pi_1 = 0\,,\qquad
\Pi_1 +\Pi_2 = I \ee and require \be \label{prtr} \Pi_1 *\chi_\mu =
\chi_\mu  * \Pi_2 =\chi_\mu \,, \qquad \Pi_2
*  \bar{\chi}^\mu =  \bar{\chi}^\mu* \Pi_1 =
\bar{\chi}^\mu\,, \ee \be \Pi_1 * \phi^A = \phi^A *\Pi_1 =
0\,,\qquad \Pi_2 * \phi^A = \phi^A *\Pi_2 =  \phi^A\,,\qquad
\{\phi^A ,\phi^B \}_* = -2\eta^{AB}\Pi_2\,. \ee As a result we have
\be
\bar{\chi}^\mu* \bar{\chi}^\nu=0  \,,\qquad \chi_\nu
* \chi_\mu \ =0\,. \ee

In addition we require \be \label{spinr1} \chi_\nu * \bar{\chi}^\mu
= -i\delta_\nu{}^\mu \Pi_1\,, \ee and \bee \label{spinr2}
\bar{\chi}^\mu*\chi_\nu \ &=& -i\gs^{-1} (M)
(\overline{\exp}{-\phi_A\gga^A})_\nu{}^\mu *\Pi_2
\\\nn &{}&\equiv -i\gs^{-1} (M)
\sum_{p=0}^{p=M+2}(-1)^{\f{p(p+1)}{2}}\f{1}{p!} \phi^{A_1}\ldots
\phi^{A_p}*\Pi_2  \gga_{A_1\ldots A_p}{}_\nu{}^\mu \,, \eee where
$\gs(M)=2^{[M/2]+1}$ is the dimension of the spinor representation,
i.e. $\mu , \nu = 1\ldots \gs (M)$ and \be \gga_{A_1\ldots A_p} =
\gga_{[A_1} \ldots \gga_{A_p]}\equiv \gga_{[[A_1} \ldots
\gga_{A_p]]}\, \ee form a basis of the  Clifford algebra.
$\overline{\exp}{-\phi_A\gga^A}$ is defined as usual exponential
with $\phi_A$ and $\gga^A$ would be anticommuting to each other,
i.e., \be \overline{\exp}{\phi_A\gga^A}=
\sum_{p=0}^{p=M+2}(-1)^{\f{p(p-1)}{2}}\f{1}{p!} \phi^{A_1}\ldots
\phi^{A_p}  \gga_{A_1\ldots A_p}{} \,, \ee so that,  taking into
account (\ref{basrc}), one gets in accordance with (\ref{clre}) \be
\phi^B* \overline{\exp}{-\phi_A\gga^A}=
\overline{\exp}{-\phi_A\gga^A}\,\gga^B\,,\qquad
 \overline{\exp}{-\phi_A\gga^A}*\phi^B = \gga^B\,
\overline{\exp}{-\phi_A\gga^A}\,. \ee The relative coefficients in
(\ref{spinr1}) and (\ref{spinr2}) are fixed by the associativity of
the spinor generating elements. (In practice it is enough to check
that the two possible ways of computing $ \bar{\chi}^\mu*\chi_\nu
*  \bar{\chi}^\nu*\chi_\mu  $ give the same result.)

 A general element of the HS superalgebra we call
$\su$ (the meaning of this notation is explained in the
end of this section) is now represented in the form
\be
\label{sel}
a =\Delta*T*\Big ( a_{11} (Y)*\Pi_1  +a_{22}(Y,\phi)*\Pi_2
 + a_{12}^\mu (Y)*\chi_\mu  +
 \bar{\chi}^\mu *a_{21\mu}(Y)  \Big )*T\,, \ee where \be T= \Pi_1
+\Pi_2*L\,, \ee $a_{11}(Y)$, $ a_{12}^\mu (Y)$ and $a_{21\mu}(Y) $
are some polynomials of $Y^A_i$, and $a_{22}(Y,\phi)$ is a
polynomial of $Y^A_i$ and $\phi^A$, such that they all  commute to
the $sp(2)$ generators $t_{ij}$.

In fact, the projectors $\Pi_1$ and $\Pi_2$ are introduced
to parametrize four blocks of a matrix which contains
elements of the bosonic algebras $\hu$ and $\huo$
  in the diagonal blocks associated with the projectors
$\Pi_1 $ and $\Pi_2 $, respectively, while odd elements of the HS
superalgera are contained in the off-diagonal blocks. The
coefficients $a_{11}$ and $a_{22}$ are assumed to be Grassmann even
(commuting) while $a_{21\mu} $ and $ a_{12}^\mu$ are Grassmann odd
(anticommuting). This convention induces the superalgebra structure
through the standard definition of field strengths with Grassmann
odd spinor gauge fields. Note that the introduced projector
structure can conveniently be described by the auxiliary Clifford
variables $ \theta*\theta = \bar{\theta}*\bar{\theta}=0 \,,\{\theta
, \bar{\theta}\}_* = 1\,,$ which have zero star commutators with all
other generating elements. Then $\Pi_1 = \theta* \bar{\theta}$,
$\Pi_2 =  \bar{\theta}*\theta $, $\chi_\nu $ contains one power of
${\theta}$ and $ \bar{\chi}^\mu $ contains one power of
$\bar{\theta}$.

By construction, the elements $ a_{12}^\mu (Y)$ of the form $
\tilde{a}_{12}^\nu (Y)*Y^A_i \gga_A{}_\nu{}^\mu$ do not contribute
to (\ref{sel}) as well as elements $a_{21\mu} $ of the form
$\gga_A{}_\nu{}^\mu Y^A_i
* \tilde{a}_{21\mu}(Y) $. As a result, representatives of the
fermionic sectors of the superalgebra can be chosen in the form \be
a_{21\mu}(Y)  =\sum_{n=0}^\infty a_\mu^{A_1\ldots A_n ,B_1\ldots
B_n} Y_{A_1}\ldots Y_{A_n}\, Y_{B_1}\ldots Y_{B_n}\,, \ee \be
a_{12}^\mu (Y)  =\sum_{n=0}^\infty  \bar{a}^{\mu\, A_1\ldots A_n
,B_1\ldots B_n} Y_{A_1}\ldots Y_{A_n}\, Y_{B_1}\ldots Y_{B_n}\,, \ee
where the spinor-tensors $a_\mu^{A_1\ldots A_n ,B_1\ldots B_n} $ and
$ \bar{a}^{\mu\, A_1\ldots A_n ,B_1\ldots B_n} $ have symmetry
properties of the two-row rectangular Young tableau with respect to
the indices $A$ and $B$ (i.e. symmetrization over any $n+1$ indices
gives zero) and satisfy the $\gamma$-transversality conditions \be
\gga_{A_1\nu}{}^\mu a_\mu^{A_1\ldots A_n ,B_1\ldots B_n} =0\,,\qquad
\bar{a}^{\mu\, A_1\ldots A_n ,B_1\ldots B_n} \gga_{A_1\,\mu}{}^\nu{}
=0\,. \ee

As a result, the gauge fields associated with the superalgebra
$\su$ consist of bosonic and fermionic 1-forms.
Bosonic gauge fields corresponding to the subalgebras
$\hu$ and $\huo$ are listed in the end of section
 \ref{Simplest bosonic higher spin algebras}.
Fermionic fields $dx^\un \go_\un{}_{\mu}^{A_1\ldots A_n ,B_1\ldots
B_n}(x) $ and $ dx^\un \bar{\go}_\un{}^{\mu\, A_1\ldots A_n
,B_1\ldots B_n} (x) $ belong to the two-row rectangular
$\gga$-transverse spinor-tensor representations of $o(d-1,2)$. These
correspond to totally symmetric half-integer spin
 massless representations of $ o(d-1,2)$.

The chiral superalgebras $\supm$ are obtained from\hfill\\ $\su$
with the aid of the projectors $\Pi_\pm$ (\ref{pipm})
\be
\label{chel}
f\in \supm : \quad f=\Pi_\pm *g *\Pi_\pm \,,\quad g\in\su\,.
\ee
For even $M$ the projection (\ref{chel}) implies chiral projection
for spinor generating elements and projects out bosonic elements which
are odd in $\phi$. For odd $M$ this condition just
implies irreducibility of the spinor representation
of the Clifford algebra.

Now let us define a family of HS superalgebras $\hunm$
with non-Abelian spin 1 subalgebras
(i.e., Chan-Paton indices). To this end we
consider the algebra of operator-valued
$(n+m)\times (n+m)$ matrices of the form (\ref{sel})
such that the
elements $a_{11}\to a_{11}{}^u{}_v$ are $n\times n$ matrices
($u,v=1,\ldots n$),
elements $a_{22}\to a_{22}{}^{u^\prime}{}_{v^\prime}$
are $m\times m$ matrices
($u^\prime,v^\prime=1,\ldots m$),
elements $a_{12}\to a_{12}{}^u{}_{v^\prime}$ are $n\times m$ matrices,
and elements $a_{21}\to a_{21}{}^{u^\prime}{}_{v}$
 are $m\times n$ matrices.

The reality conditions which single out the appropriate real HS
superalgebra $\hunm$ are \be (a_{11}(Y))^\dagger {}^u{}_v = -
a_{11}(Y){}^u{}_v\,,\qquad (a_{22}(Y))^\dagger
{}^{u^\prime}{}_{v^\prime} = -
a_{22}(Y){}^{u^\prime}{}_{v^\prime}\,, \ee and \be \label{nmreal} (
a_{12}^\mu (Y))^\dagger{}^{u^\prime}{}_{v}
 =a_{21\mu}{}^{u^\prime}{}_{v}(Y)  \,,\qquad
(\chi_\mu  )^\dagger = i  \bar{\chi}^\mu \,, \ee where $\dagger$
denotes usual matrix Hermitian conjugation along with the
conjugation of the generating elements $Y^A_i$ and $\phi^A$
according to (\ref{defr}) and (\ref{phidag}). It is easy to see that
such conditions indeed single out a real form of the complex Lie
superalgebra resulting from the original associative algebra with
elements (\ref{sel}) by (anti)commutators as a product law. Note
that $\hunm$ contains $u(n)\oplus u(m)$ as a finite dimensional
subalgebra. The labels 0 and 1 in this notation indicate how many
Clifford elements taking values in the vector representation of
$o(M,2)$ appear in the respective diagonal blocks.

Analogously to the case of $4d$ HS algebras \cite{KV1}, the algebras
$\hunm$ admit truncations by an antiautomorphism $\rho$ of the
original associative algebra to the algebras $\honm$ and $\huspnm$
with $o(n)\oplus o(m)$ and $usp(n)\oplus usp(m)$ (here $n$ and $m$
are even) as finite-dimensional Yang-Mills (spin 1) subalgebras. The
truncation condition is \be \rho (a) = - i^{\pi (a)} a\,, \ee where
$\pi(a)= 0$$(1)$ for (even) odd elements of the superalgerba. The
action of $\rho$ on the matrix indices is defined as usual \be
\rho({a_u{}^v })= - \rho^{vp} a_p{}^q \rho_{q u}\,,\quad
\rho({a^\prime_{u^\prime}{}^{v^\prime}) }= - \rho^{ v^\prime
p^\prime } a^\prime_{p^\prime}{}^{q^\prime} \rho_{q^\prime
u^\prime}\,, \ee where the bilinear forms $\rho_{uv}$ and
$\rho_{u^\prime v^\prime}$ are non-degenerate and either both
symmetric (the case of $\honm$) or antisymmetric (the case of
$\huspnm$). The action of $\rho$ on the generating elements is
defined by the relations \be \rho (Y^A_j ) = i Y^A_j \qquad \rho
(\phi^A ) = -\phi^A\,,\qquad \rho (\chi_\nu{}^u{}_{u^\prime} )=
\bar{\chi}^\mu {}^{v^\prime}{}_v \rho^{uv} \rho_{v^\prime u^\prime}
C_{\mu\nu}\,, \ee where $C_{\mu\nu}$ is the charge conjugation
matrix which represents the Clifford algebra antiautomorphism
$\rho(\phi^A )$ in the chosen representation of $\gamma$ matrices,
i.e. \be \gamma^A{}_\nu{}^\mu
=-C^{\mu\sigma}\gamma^A{}_\sigma{}^\eta C_{\eta\nu} \,, \qquad
C^{\mu\sigma}C_{\nu\sigma}  =\delta^\mu_\nu \,. \ee

The chiral HS algebras $\hunmpm$ , $\honmpm$  and $\huspnmpm$
are obtained from $\hunm$,\hfill\\ $\honm$ and $\huspnmpm$ by the projection
(\ref{chel}), taking into account that $\rho (\Gamma) = \Gamma$ and,
therefore, $\rho (\Pi_\pm) = \Pi_\pm$.

Finally let us note that
the algebras $\hunmpq$ with $u$ and $v$
copies of fermions in the upper and lower blocks are
 likely to be relevant to the analysis of HS gauge theories
with mixed symmetry HS gauge fields. A number of copies of bosonic
oscillators, which are assumed to be the same in the
upper and lower blocks, can also be enlarged to the case of
$hu(n,m,|(u,v,p)\!\!:\!\![M,2])$.
In this notation
$\hu\sim hu(1,0|(0,0,2)\!\!:\!\![M,2])$ and
$\huo\sim hu(1,0|(1,0,2)\!\!:\!\![M,2])$.
More generally, (super)algebras
with $u$ fermions and $p$ bosons in the upper block and
$v$ fermions and $q$ bosons in the lower block (with all bosons and
fermions taking values in the vector representation of $o(s,t)$)
are denoted  $hu(n,m,|(u,p;v,q)\!\!:\!\![s,t])$ (and similarly
for the orthogonal ($ho$) and symplectic $(husp)$
series). Note that these algebras
are Lie superalgebras when $u+v$ is odd.

\section{Spinorial realizations in lower dimensions}
\label{Spinorial realization}

It is instructive to compare the HS superalgebras introduced in this
paper with the  HS superalgebras in lower space-time dimensions,
defined earlier in terms of spinor oscillator algebras. For example,
the simplest $AdS_4$ superalgebra $hu(1,1|4)$ (in notations of
\cite{KV1}) was realized \cite{Fort2,FVA} as the algebra of
functions $f(y,K)$ of the spinor oscillators $y_\mu$ and Klein
operator $K$ satisfying relations \be \label{spir} [y_\mu , y_\nu
]_* = 2i C_{\mu\nu}\,,\qquad K*y_\mu = -y_\mu *K\,, \qquad K^2 =
1\,, \ee where $C_{\mu\nu}=-C_{\nu\mu}$ is the $4d$ antisymmetric
charge conjugation matrix. The spinorial generating elements $y_\mu$
are not subject to any further restrictions.

One can try to identify the generating elements $\chi_\mu $ and $
\bar{\chi}^\mu$ with such independent spinorial generating elements.
This does not work however for the general case because of the
projectors $T$ in (\ref{sel}) which induce a nontrivial dependence
on the bosonic and fermionic oscillators $\phi^A$ and $Y^A_i$ into
$\chi_\mu $ and $ \bar{\chi}^\mu$-dependent terms. To see what
happens let us start with the case of general $M$.

First one observes that the operator
\be
\label{K}
K=\Pi_1 - \Pi_2 \,,\qquad
K*K = 1
\ee
behaves just as the Klein operator. {}From (\ref{prtr})
it follows that it anticommutes with fermions.
Let $C_{\mu\nu}$ be the charge conjugation matrix which is either
symmetric or antisymmetric depending on $M$,
\be
C_{\mu\nu}=(-1)^{\gga(M)}C_{\nu\mu}\,.
\ee

Let us set \be y_\mu = 2 T*\Big ( \chi_\mu  + \bar{\chi}^\nu
C_{\nu\mu}\Big )*T\,. \ee (To simplify formulae, in this section we
discard the operator $\Delta$ which is an overall factor in all
expressions.) {}From  (\ref{K}) it follows that \be K*y_\mu = 2
T*\Big ( \chi_\mu  - \bar{\chi}^\nu C_{\nu\mu}\Big )*T\,. \ee As a
result the set of elements $y_\mu$ and $K* y_\mu$ is as good  as the
original set of elements $ \chi_\mu  *L$ and $L* \bar{\chi}^\nu$ in
the decomposition $(\ref{sel})$. {}From the defining relations
(\ref{spinr1}) and (\ref{spinr2}) we obtain \be \label{1-K} \Pi_1 *
y_\mu *y^\nu = 4 \Pi_1* \chi_\mu *L* \bar{\chi}^\nu\,, \ee \be
\Pi_2* y^\nu *y_\mu = 4 \Pi_2*L* \bar{\chi}^\nu* \chi_\mu  *L\,. \ee
These relations can be treated as defining relationships on spinor
generating elements of the algebra. By means of (\ref{ldef}),
(\ref{clre}), (\ref{spinr1}), (\ref{spinr2}) they get an equivalent
form \be \label{p1} \Pi_1 * y_\mu *y^\nu =4i \Pi_1* \Big (
\f{1}{4}(M-2)\delta_\mu^\nu +\f{i}{8} \gga^{AB}{}_\mu{}^\nu
Y_{Aj}Y_B^j \Big )\,, \ee \be \label{p2} \Pi_2* y^\nu *y_\mu
=-4i\gs^{-1} (M) \Pi_2*L*
 \sum_{p=0}^{p=M+2}(-1)^{\f{p(p+1)}{2}}\f{1}{p!}
\phi^{A_1}\ldots \phi^{A_p}*L  \gga_{A_1\ldots A_p}{}_\nu{}^\mu \,.
\ee
The resulting expressions depend nontrivially on $\phi^A$ and $Y^A_i$
that means in particular that, generically,
 the anticommutator of the spinorial
element $y_\mu$ gives rise to HS generators. Since $y_\mu$ and
$Ky_\mu$ are the only candidates for usual supercharges, the expressions
(\ref{p1}) and (\ref{p2}) indicate that
the HS superalgebras under consideration possess
no finite dimensional conventional SUSY subsuperalgebras
for general $M$.

Let us now turn to the case of $M=3$ which corresponds to $AdS_4$.
According to (\ref{oddec}), the irreducibility in the case of odd
$M$ implies that one has to work with the algebras
$hu_\pm(1,1|(0,1,2)\!\!:\!\![M,2])$ (\ref{chel}) which implies that
independent elements on the right hand side of (\ref{1-K}) are even
in the Clifford elements $\phi^A$. For $M=3$ the terms of zeroth,
second and fourth order can appear. The key point is that, according
to Lemma 3.1 the fourth order terms in $\phi$ do not contribute
(factor out). As a result, only the terms containing
$\delta_\nu{}^\mu$ and $\gga_{AB}{}_\nu{}^\mu$ survive. With lowered
index $\mu$ these have opposite symmetry types. As a result, one
obtains that the spinor variables $y_\mu$ do indeed  satisfy the
Heisenberg commutation relations (\ref{spir}).  Clearly, this is not
true for generic $M$ because fourth order and higher order terms in
$\phi$ will contribute to the right hand side of the commutator of
the spinor generating elements.

The anticommutator has the form \be \label{sym} \{y_\mu , y_\nu \}_*
= (\ga \Pi_1 +\gb \Pi_2 *L)*T_{AB} \gga^{AB}{}_{\mu\nu} \,, \ee with
some nonzero $\ga$ and $\gb$, where $T_{AB}$ is the $o(M,2)$
generator (\ref{TS}). Now one observes that the terms on the right
hand side of (\ref{sym}) parametrize all $osp(1,2)$ invariant
bilinear combinations of oscillators $Y_{Ai}$ and $\phi_A$ (in fact,
this is the manifestation of the isomorphism $sp(4)\sim o(3,2)$).
Taking into account that $sp(2)$ invariant polynomials of $Y_{Ai}$
and $osp(1,2)$ invariant polynomials in $Y_{Ai}$ and $\phi_A$ are
star polynomials of the invariant bilinears $T^{AB}$, one concludes
that  $y_\nu$ and $K$ form an equivalent set of generating elements
for the $M=3$ HS superalgebra, i.e. one can forget about the
generating elements $Y_{Ai}$ and $\phi_A$ in the case of $M=3$, thus
arriving at the purely spinorial realization of the $4d$ HS algebras
originally introduced in \cite{Fort2,FVA} and denoted $hu(1,1|4)$ in
\cite{KV1}.

The analysis of $AdS_3$ \cite{bl,V3} HS algebras is analogous to
that of the $4d$ case: because the terms of fourth order in $\phi^A$
do not appear, the spinor generating elements $y_\mu$ along with the
Klein operator can be chosen as independent generating elements thus
establishing isomorphism with the original spinor realization. Since
the $AdS_3$ algebra is semisimple $o(2,2)\sim o(2,1)\oplus o(2,1)$,
the corresponding HS extensions are also direct sums of the $M=1$
algebras. This fact manifests itself in the isomorphism
$hu(1|2\!\!:\!\![2,2])\sim hu(1|2\!\!:\!\![1,2])\oplus
hu(1|2\!\!:\!\![1,2])$ which is not hard to prove by observing that
any length $h$ rectangular two-row Young tableau of $o(2,2)$
decomposes into the direct sum of one selfdual and one anti-selfdual
length $h$ rectangular two-row Young tableaux\footnote{As one can
easily see, imposing opposite (anti)selfduality conditions on two
pairs of tensor indices associated with different columns in a
$o(2,2)$ Young tableau gives zero.}, each forming a
$o(2,1)$--module.

The isomorphisms between spinorial and vectorial realizations of
$AdS_3$ and $AdS_4$ HS superalgebras extend to their matrix
extensions $hu(n,m|2k)$, $ho(n,m|2k)$ and $husp(n,m|2k)$ considered
originally in \cite{KV1} for the spinorial realization and those
considered in the end of the previous section for the vectorial
realization. Namely, \be hu(n,m|4)\sim
hu_\pm(n,m|(0,1,2)\!:\![3,2])\,,\qquad hu(n,m|2)\sim
hu_\pm(n,m|(0,1,2)\!:\![1,2]) \,,\ee \be ho(n,m|4)\sim
ho_\pm(n,m|(0,1,2)\!:\![3,2])\,,\qquad ho(n,m|2)\sim
ho_\pm(n,m|(0,1,2)\!:\![1,2]) \,,\ee \be husp(n,m|4)\sim
husp_\pm(n,m|(0,1,2)\!:\![3,2])\,,\quad husp(n,m|2)\sim
husp_\pm(n,m|(0,1,2)\!:\![1,2])\,. \ee

{}From here it follows that usual $AdS_4$ superalgebras $osp(N,4)$
are subalgebras of the HS superalgebras with the parameters $n$ and
$m$ high enough. Namely, one has for even $N$ \cite{KV1} \be
osp(2p,2k)\subset hu(2^{p},2^{p}|2k)\,, \ee
\be osp(8p,2k)\subset
ho(2^{4p-1},2^{4p-1}|2k)\,,
\ee
\be osp(8p+4,2k)\subset
husp(2^{4p+1},2^{4p+1}|2k)\,.
\ee

Analogously, it follows also that $AdS_3$ SUSY algebras
$osp(N_+,2)\oplus osp(N_-,2)$ are subalgebras of the appropriate
$3d$ HS algebras of the form $h\ldots (n_+,m_+|2)\oplus h\ldots
(n_-,m_-|2)$ where dots denote one of the three possible types of
the algebra (unitary, orthogonal or unitary symplectic) and $n_\pm$
are appropriate $N_\pm$--dependent powers of two.

The case of $AdS_5$ corresponds to $M=4$. Here one takes the chiral
HS algebra $hu_\pm(1,1|(0,1,2)\!\!:\!\![4,2])$. Again, in this case
effectively only zero-order and second-order combinations of
Clifford elements appear and it is possible to establish the
isomorphism between the spinorial realization of the $AdS_5$
(equivalently $4d$ conformal) HS algebra and that given in this
paper. Note that the spinorial realization of the $AdS_5$ HS algebra
also includes some reduction procedure which assumes a restriction
to the  centralizer of some operator $N$ followed by the
factorization of the elements proportional to $N$ (see
\cite{FLA,SS5,BHS}). This does not allow one to express the spinor
generating elements of the original $5d$ spinorial construction
directly in terms of the generating elements of this paper. In fact
it is this property which makes it impossible to extend this
isomorphism to the case of $5d$ HS superalgebras which contain
higher $N$--extended finite dimensional subsuperalgebras.

A simplest way to see this is by using the facts shown in the rest
of this paper that the HS superalgebras considered here act on
massless scalar and spinor singletons (i.e., boundary conformal
fields). On the other hand, the conformal realization \cite{BHS} of
the spinorial $AdS_5$ HS algebras of \cite{FLA,SS5} deals with
various $4d$  massless boundary supermultiplets which contain spins
$s\geq1$ for $N>2$ . Thus, the maximal conventional $N$--extended
$5d$ supersymmetry compatible with the HS algebras of the type
considered in this paper is that with $N=2$ associated with the
boundary massless hypermultiplet. In notation of \cite{BHS} the
corresponding HS algebra is $husp_0(2,2|8)$ while its finite
dimensional subsuperalgebra is $su(2,2|2)$. We therefore expect
that, $su(2,2|2)\subset husp_\pm(2,2|(0,1,2)\!:\![4,2])$. The
algebra $psu(2,2|4)$ associated with the $N=4$ SYM supermultiplet is
not a subalgebra of the HS algebras considered in this paper just
because the $N=4$ SYM supermultiplet contains spin one massless
states absent in the scalar-spinor singleton realization of the HS
algebras of this paper.

For analogous reasons we expect that the purely bosonic $7d$ HS
algebra of \cite{SS7} is isomorphic to $hu_\pm(1|2\!\!:\!\![6,2])$
while the $M=6$ HS superalgebras considered in this paper are all
different from those discussed in \cite{SS} (which contain the
finite dimensional subsuperalgebra denoted $osp(8^*|4)$ in
\cite{SS}) because the latter are associated with the tensor
singletons absent in the construction of this paper.

For $M>6$, terms with higher combinations of the fermionic
oscillators and $\gamma$ matrices appear in the defining relations
for spinorial elements $y_\mu$ that complicates the spinorial
realization of the HS algebras of this paper. Note, however, that
the superalgebras with unrestricted spinorial elements suggested in
\cite{vasfer}
 may still be relevant for the description of HS theories
with mixed symmetry HS gauge fields in higher dimensions.

\section{Scalar conformal module}
\label{rac} According to notations of \cite{FF}, {\it Rac} is the
unitary representation of $o(3,2)$ realized by a $3d$ conformal
massless scalar field. The global conformal HS symmetry of a
massless scalar in $M$-dimensions $\hu$ (more precisely, its
complexification) was originally introduced by Eastwood in
\cite{East}. For our purpose it is most convenient to use its
realization in terms of bosonic oscillators as explained in section
\ref{Simplest bosonic higher spin algebras}.

Let us  introduce  mutually conjugated oscillators \be a^A = \half
(Y^A_1 - i Y^A_2 )\,,\qquad \bar{a}^A = \half (Y^A_1 + i Y^A_2 )\,,
\qquad (a^A )^\dagger = \bar{a}^A\,, \ee which satisfy the
commutation relations \be [a^A , \bar{a}^B ] = -\eta^{AB}\,,\qquad
[a^A , {a}^B ] =0 \,,\qquad [\bar{a}^A , \bar{a}^B ] = 0\,. \ee For
the space-like values of $A=a=1\ldots M$ with $\eta^{ab} =
-\delta^{ab}$ these are normal commutation relations for creation
and annihilation operators. For the time-like directions $A=0,M+1$
it is convenient to introduce the set of oscillators \be \bar{\ga} =
a^0 +i a^{M+1} \,,\qquad \bar{\gb} = a^0 -i a^{M+1} \,,\qquad {\ga}
= \bar{a}^0 -i \bar{a}^{M+1} \,,\qquad {\gb} = \bar{a}^0 +i
\bar{a}^{M+1}  \ee $ (\ga^\dagger = \bar{\ga}\,,\quad \gb^\dagger =
\bar{\gb})\,, $ which have the nonzero commutation relations \be
[\ga , \bar{\ga} ] =2\,,\qquad [\gb , \bar{\gb} ] =2\,. \ee

The generators of the algebra $o(M,2)$ are
\be
T^{AB}= - (\bar{a}^A a^B - \bar{a}^B a^A )\,.
\ee
The $AdS_{M+1}$ energy operator (\ref{ener}) is
\be
E =  \half (\bar{\ga}\ga - \bar{\gb}\gb )\,.
\ee
The noncompact  generators of $o(M,2)$ are
\be
\label{tpm}
T^{+a} =
-i ( \gb a^a - \bar{\ga} \bar{a}^a)\,,\qquad
T^{-a} =
i(\bar{\gb} \bar{a}^a - {\ga} {a}^a)\,.
\ee

Let us now introduce the  Fock module $U$  of states
$
|\Psi \rangle = \psi (\bar{a},\bar{\ga},\bar{\gb} ) |0\rangle \,
$
induced from the vacuum state
\be
a^a |0\rangle =0\,,\qquad
\ga |0\rangle =0\,,\qquad
\gb |0\rangle =0\,.
\ee
$U$ is endowed with the positive definite norm with respect to
which $a^A$ is conjugated to $\bar{a}^A$, i.e.  the conjugated
vacuum $\langle 0|$ is defined by
\be
\langle 0| \bar{a}^a =0\,,\qquad
\langle 0| \bar{\ga} =0\,,\qquad
\langle 0| \bar{\gb} =0\,.
\ee

Let us note that the vacuum vector $|0\rangle$ is not annihilated by
$T^{-a}$ in contrast with the standard singleton construction in
terms of spinorial oscillators used for the description of singleton
and doubleton modules  \cite{Gun1}. We consider the singleton
submodule $S\subset U$ spanned by the $sp(2)$ invariant states
satisfying $t_{ij}|\Psi \rangle=0$. These conditions are equivalent
to \be \label{spc} \tau^- |\Psi \rangle= \tau^+ |\Psi \rangle=\tau^0
|\Psi \rangle= 0\,, \ee where \be \tau^- = a_a a^a
+\bar{\ga}\bar{\gb}\,,\qquad \tau^+ = \bar{a}_a \bar{a}^a
+{\ga}{\gb}\,,\qquad \tau^0 = \{a_a, \bar{a}^a\} +\half \{\ga,
\bar{\ga}\} +\half\{\gb, \bar{\gb}\} \, \ee (note that $a_a b^a =
a_a b_b \eta^{ab}$ with $\eta^{ab} = - \delta^{ab}$.) By its
definition, $S$ forms a $\hu$--module. Let us note that the
conditions (\ref{spc}) are consistent because the $o(M,2)$ invariant
metric has signature $(M,2)$. In the case of Euclidean signature,
for example, the conditions (\ref{spc}) associated with $\tau_0$ and
$\tau^+$ would allow no solution at all in a unitary module. Note
that, although some details of the realization of appropriate
modules are different, our construction is a version of that of the
two-time formalism developed by Marnelius and Nilsson \cite{mar} and
Bars and collaborators \cite{Bars}, in which the $sp(2)$ invariance
condition also plays the key role.

The generic solution of the conditions (\ref{spc}) is
\be
|\Psi (\bar{a},\bar{\ga},\bar{\gb} )\rangle =
|\Psi^+ (\bar{a},\bar{\ga},\bar{\gb} )\rangle +
|\Psi^- (\bar{a},\bar{\ga},\bar{\gb} )\rangle\,,
\ee
\be
|\Psi^\pm (\bar{a},\bar{\ga},\bar{\gb} )\rangle =
\psi^\pm (\bar{a},\bar{\ga},\bar{\gb} ) |0\rangle \,,
\ee
where
\be
\label{+}
\psi^+ (\bar{a},\bar{\ga},\bar{\gb} ) =
\sum_{p,n=0}^\infty \f{(-1)^p}{2^{2p} p! (p+n+\half M-1)!}
(\bar{a}_a \bar{a}^a)^p
\bar{\ga}^{p+n+\half M-1}\bar{\gb}^p \psi^+_n(\bar{a})\,,
\ee
\be
\label{-}
\psi^- (\bar{a},\bar{\ga},\bar{\gb} ) =
\sum_{p,n=0}^\infty \f{(-1)^p}{2^{2p} p! (p+n+\half M-1)!}
(\bar{a}_a \bar{a}^a)^p
\bar{\gb}^{p+n+\half M-1}\bar{\ga}^p \psi^-_n(\bar{a})
\ee
and $\psi^\pm_n(\bar{a})$ are arbitrary degree $n$
harmonic polynomials, i.e.
\be
\label{harm}
\psi^\pm_n(\bar{a})=\psi^\pm_{a_{1}\ldots a_{n}}\bar{a}^{a_1} \ldots
\bar{a}^{a_n} \,,\qquad \psi^\pm_{bc a_3\ldots a_n}\eta^{bc} =0\,.
\ee

Let us note that the Fock space realization of the modules
$|\Psi^\pm \rangle = \psi^\pm |0\rangle $ is literally valid for
even $M$ when all powers of oscillators are integer. Nevertheless,
the modules $|\Psi^\pm \rangle $ are well-defined for odd $M$ as
well. Actually, although powers of one of the oscillators
$\bar{\ga}$ or $\bar{\gb}$ are half-integer for odd $M$,  the
modules $|\Psi^\pm \rangle $ are semi-infinite because the powers
of the another oscillator $(\bar{\gb}$ or $\bar{\ga})$ are
nonnegative integers. We therefore will use the formulae (\ref{+})
and (\ref{-}) for all $M$.

The expressions (\ref{+}) and (\ref{-}) admit the
following useful form
\be
\label{uf+}
|\Psi^\pm \rangle = \bar{{\cal P}}^\pm \psi^\pm (\bar{a} ) |0\rangle\,,
\ee
where $\psi^\pm (\bar{a} ) $ are the harmonic polynomials
(\ref{harm}) and
\be
\label{uf-}
\bar{{\cal P}}^+ = \oint d \mu
\exp{(\mu^{-1}\bar{\ga}-\f{1}{4}{\mu \bar{a}^a \bar{a}_a \bar{\beta}}
)}\mu^{-\half\{a^b, \bar{a}_b\}-2}\,,
\ee
\be
\bar{{\cal P}}^- = \oint d \mu
\exp{(  \mu^{-1}\bar{\gb} -\f{1}{4}\mu \bar{a}^a \bar{a}_a \bar{\ga})
}\mu^{-\half\{a^b, \bar{a}_b\}-2}\,,
\ee
where the integral is defined by
\be
\oint d\mu \mu^{-p} = \delta (p-1) \,,\qquad
\delta(p) = 1 (0)\,,\qquad p=0 (\neq 0)\,.
\ee
Using this representation it is easy to check that (\ref{spc}) is
true. Let us note that the operators $\bar{{\cal P}}^\pm$
cannot be  expanded in power series of the oscillators since they
effectively contain terms of the type $\ga^{-\half\{a^b, \bar{a}_b\}-1}$
or $\gb^{-\half\{a^b, \bar{a}_b\}-1}$.

The Fock space  norm of the states $|\Psi^\pm \rangle$ diverges.
Indeed it is easy to see that $\langle \bar{\Psi}^+ |\Psi^+\rangle
\sim \sum_{p=0}^\infty 1 $ using that \be \label{prop} \langle
\bar{\Psi}_n | (a_a a^a )^p (\bar{a}_a \bar{a}^a )^p | \bar\Psi_n
\rangle =2^{2p} \f{p! (\half M +p+n -1)!}{(\half M +n -1)!}\,, \ee
where $n$ refers to a power of $\psi^+_n$ (\ref{harm}). This fact
is neither occasional nor problematic, being a manifestation of
the standard inner product problem with the Dirac quantization
prescription for  first-class constraints. Indeed, here the
``first class constraints"  (\ref{spc}) form the Lie algebra
$sp(2)$. For normalizable states  annihilated by the first-class
constraints $\tau^i |\Psi_{1,2} \rangle =0$ one obtains \be
 \langle \bar{\Psi}_1|(\tau^i  \chi_1 +\chi_2 \tau^i
) |\Psi_2\rangle=0 \,
\ee
for any $\chi_{1,2} $. Choosing
appropriately $\chi_{1,2}$ one finds a contradiction with the
assumption that the corresponding matrix element is finite. For
example, choosing $\chi_1 =-\chi_2 = \ga\gb$ one finds for
$\tau^-$ that all matrix elements of the manifestly
positive-definite operator $\{\ga , \bar{\ga}\} + \{\gb ,
\bar{\gb}\}$ must vanish, that means that they cannot be finite.
Note that this phenomenon has the same origin as the fact of
non-existence of $\Delta^2$ for a projection-like operator
$\Delta$ of section \ref{Projection technics}. A way out is also
analogous.

A well-defined scalar product $\lla|\,| \rr$ for the $sp(2)$
invariant module must break down the $sp(2)$ invariance of the
Fock scalar product. This is achieved by redefining the scalar
product in the form
\be
\label{renorm}
\lla \bar{\Psi}_1 |\Psi_2
\rr = \langle \bar{\Psi}_1 |G|\Psi_2 \rangle\,,
\ee
where $G$ is
some operator which does not commute with the $sp(2)$ generators.
We demand the scalar product $\lla \bar{\Psi}_1 |\Psi_2 \rr$  to
induce an $o(M,2)$ invariant norm on the $sp(2)$ invariant states.
This can be achieved by choosing such $G$ that \be [T^{AB} , G]=
[\tau^{ij} , X^{AB}_{ij}] \ee for some $X^{AB}_{ij}$.

The appropriate operators $G^\pm$ for the modules $|\Psi^\pm\rangle$
are
\be
\label{bpm}
G^+ = ( \bar{\ga}\ga +2)^{-1}  F^+ \,,\qquad
G^- = ( \bar{\gb}\gb +2)^{-1}  F^-\,,
\ee
where $F^+$ and $F^- $ are the Fock projectors
for the oscillators $\gb, \bgb$ and $\ga,\bga$
respectively,
\be
\gb F^+ =0\,,\qquad
F^+\bgb =0\,,\qquad
[\ga\,, F^+ ] =0\,,\qquad
[\bga\,, F^+ ] =0\,,\qquad
\ee
\be
\ga F^- =0\,,\qquad
F^-\bga =0\,,\qquad
[\gb\,, F^- ] =0\,,\qquad
[\bgb\,, F^- ] =0\,,\qquad
\ee
\be
[a_a\,, F^\pm ] =0\,,\qquad
[\ba_a \,, F^\pm ] =0\,,\qquad
\ee
which are normalized so that
\be
\langle 0| F^\pm |0\rangle =1\,.
\ee
The Fock projectors $F^\pm$ can be realized as
\be
F^+ = \sum_{n,m=0}^\infty \f{(-1)^n}{2^m n!m!}
\bar{a}_{a_1} \ldots \bar{a}_{a_n}(\bar{\ga})^m
|0\rangle \langle 0| a^{a_1}\ldots a^{a_n} (\ga)^m\,,
\ee
\be
F^- = \sum_{nm} \f{(-1)^n}{2^m n!m!}
\bar{a}_{a_1} \ldots \bar{a}_{a_n}(\bar{\gb})^m
|0\rangle \langle 0| a^{a_1}\ldots a^{a_n} (\gb)^m\,.
\ee

The scalar products \be \label{inf} \lla \bar{\Psi}_1 |\Psi_2
\rr_\pm = \langle \bar{\Psi}^\pm _1 |G^\pm |\Psi_2^\pm \rangle\,
\ee are $o(M,2)$ invariant. This property is manifest for the
generators $T^{ab}$ and $E$, respectively. For the noncompact
generators $T^{\pm a}$ this is also true because
\be
[T^{+ a}
,G^+ ]= i [\tau^+ , \bar{\ga} a^a
(\bar{\ga}\ga\ga\bar{\ga})^{-1} F^+ ]\,,\qquad [T^{- a} ,G^+ ]= i
[\tau^- ,
\ga \bar{a}^a (\bar{\ga}\ga\ga\bar{\ga})^{-1} F^+
]\,,
\ee
\be [T^{+ a} ,G^- ]=  i [\tau^+ , \bar{\gb} a^a
(\bar{\gb}\gb\gb\bar{\gb})^{-1} F^- ]\,,\qquad [T^{- a} ,G^- ]= i
[\tau^- ,      \gb \bar{a}^a (\bar{\gb}\gb\gb\bar{\gb})^{-1} F^-
]\,,
\ee
as can be seen using (\ref{tpm}). Note that the
operators $\ga\bar{\ga}$ and $\bar{\ga}\ga\ga\bar{\ga}$ are
positive definite and therefore their inverse operators are well
defined.

In terms of components (\ref{harm}), the scalar product
(\ref{renorm}) gets the manifestly positive-definite form \be \lla
\bar{\Psi}_1^\pm |\Psi_2^\pm \rr = \sum_{n=0}^\infty \f{2^{n+\half
M -2}n!}{(n+\half M )!} \bar{\psi}_n^\pm{}_{a_1 \ldots a_n}
{\psi}_n^\pm{}^{a_1 \ldots a_n}\,. \ee As a result, $|\Psi^\pm
\rangle $ form unitary $o(M,2)$-modules.

The modules $|\Psi^+ \rangle $ and $|\Psi^-\rangle $ have, respectively,
positive and negative energies
\be
E^+_n =  n+\half M-1\qquad n=0,1,2\ldots\,,
\ee
\be
E^-_n =  - n-\half M +1\qquad n=0,1,2\ldots\,.
\ee
 The lowest energy of $|\Psi^+ \rangle $ therefore is
\be \label{E0} E^+_0 = \half M-1\,. \ee This is the correct value
for the conformal scalar in $M$ dimensions \footnote{Note that the
operator  $E$  here is some combination of $P^0$ and $K^0$ in the
conformal algebra. The value (\ref{E0}) coincides with the scaling
dimension for the complex equivalent scalar field conformal
module.}. {}From (\ref{tpm}) it follows that the lowest energy state
of $|\Psi^+ \rangle $ annihilated by $T^{-a}$ is that with the
$\bar{a}$ independent part of $\psi^+_0 (\bar{a})$ in (\ref{+}),
which is $o(M)$ singlet. {}From (\ref{cas}) one finds that the value
of the $o(M,2)$ Casimir operator for the module $|\Psi^+\rangle $ is
(see also \cite{Bars}) \be C_2 = -\f{1}{4} (M^2 -4 )\,. \ee Taking
into account that $|\Psi^+\rangle $ is $sp(2)$ singlet, this is in
agreement with (\ref{cascas}).

Let us note that representation of the generators $T^{-a}$ and $E$
on the modes $\psi_n (\bar{a})$ has the following simple form \be
\label{t-} T^{-a} |\Psi^+ (\psi (\bar{a}) )\rangle = |\Psi^+ \left (
i\f{\partial}{\partial \bar{a}_a } \psi (\bar{a}) \right )\rangle\,,
\ee \be \label{e} E |\Psi^+ (\psi (\bar{a}))\rangle = |\Psi^+ \left(
(\bar{a}_a \f{\partial}{\partial \bar{a}_a} +\half M -1) \psi(
\bar{a}) \right)\rangle\,. \ee

\section{ Spinor conformal module}
\label{di}

According to notations of \cite{FF}, {\it Di} is the unitary
representation of $o(3,2)$ realized by $3d$ conformal
massless spinor field.
The global HS symmetry algebra which is the $M$ dimensional
conformal HS symmetry of a massless spinor is $\huo$.

Consider the Fock module generated from the vacuum state
satisfying
\bee
a^a |0\rangle_{\pm \ga} &=&0\,,\qquad
\ga |0\rangle_{\pm\ga}  =0\,,\qquad
\gb |0\rangle_{\pm\ga} =0\,,\qquad \phi |0\rangle_{\pm\ga} =0\,,\\
\phi_a |0\rangle_{\pm\ga} &=&  |0\rangle_{\pm\gb} \gamma_a{}^\gb{}_\ga
\,,\qquad
 |0\rangle_{\pm\gb}\gga^\gb{}_\ga    =\pm |0\rangle_{\pm\ga}\,,
\label{chir}
\eee
where
\be
\bar{\phi} = \half (\phi_0 -i\phi_{M+1} )\,,\quad
{\phi} = \half (\phi_0 +i\phi_{M+1} )\,,\quad \phi^2 =0\,,\quad
{\bar{\phi}}^2 =0\,,\quad \{\bar{\phi} , \phi\} = -1\,,
\ee
and the $\gamma$-matrices
form some representation of the Clifford algebra $C_M$ associated
with the compact algebra $o(M)\subset o(M,2)$
\be
\{\gga_a \,,\gga_b \} = -2\eta_{ab}\,,
\qquad \gga = (i)^{\half M(M-1)}\gga_1 \ldots \gga_M\,.
\ee
In other words
$|0\rangle_{\pm \ga}  $ forms a spinor representation of $o(M)$.
Here $\ga$ is the $o(M)$ spinor index while $\pm$ distinguishes
between left and right spinors. A general element
of the Fock module is
\be
\label{Psir}
|\Psi \rangle_\pm = \sum_{m,n}^\infty \sum_{s=0}^1 A(m,n,s)
(\bar{a}_a \phi^a)^m \bar{\ga}^{r(m,n,s)}\bar{\gb}^{q(m,n,s)}\bar{\phi}^s
|\psi_n (\bar{a} ) \rangle_\pm\,,
\ee
where
\be
\phi^a a_a
|\psi_n (\bar{a} )\rangle_\pm =0\,,\qquad \bar{a}^a a_a
|\psi_n (\bar{a} )\rangle_\pm  =-n |\psi_n  (\bar{a}) \rangle_\pm\,,
\ee
i.e.
\be
|\psi_n (\bar{a} ) \rangle_\pm = \psi^{a_1 \ldots a_n \,\ga}
\bar{a}_{a_1} \ldots \bar{a}_{a_n} |0\rangle_{\pm \ga}
\ee
is the generating function for rank $n$ totally symmetric
tensor-spinors $\psi^{a_1 \ldots a_n \,\ga}$
satisfying the $\gamma$-transversality condition
\be
\gamma_{a_1}{}^\ga{}_\gb\psi^{a_1 \ldots a_n \,\gb}=0\,.
\ee

The supergenerators of $osp(1,2)$ are \be Q= a^a \phi_a
+\bar{\ga}\bar{\phi} + \bar{\gb}{\phi}\,,\qquad \bar{Q}=
{\bar{a}}^a \phi_a +{\ga}{\phi} + {\gb}\bar{\phi}\,. \ee Imposing
the $osp(1,2)$ invariance condition
\be
Q |\Psi
\rangle_\pm=0\,,\qquad \bar{Q} |\Psi \rangle_\pm=0
\ee
we single out a $\huo$-module. This leads to
a set of conditions on the parameters of (\ref{Psir}) which admit
the following general solution \be q(m,n,s) = \half \Big (m-s+(2n
+M-1)|m+s+1|_2 \Big )\,,
\ee
\be
 r(m,n,s) = \half \Big (m+s-1+(2n
+M-1)|m+s|_2 \Big )\,, \ee \be A(m,n,s) =  (-1)^{\half m(m+1)}
\f{A(n, |m+s|_2 )}{(m-|m|_2 )!! (m+2n+M-1 -|m+1|_2 )!!}\,, \ee
where $A(n, |m+s|_2 )$ are arbitrary coefficients and we use
notations \be |2k|_2 =0\,,\qquad |2k+1|_2 =1\,.
\ee
The ambiguity
in the coefficients $A(n, |m+s|_2 )$ manifests the freedom of
normalization of the spinor-tensors $|\psi_n\rangle_\pm$ (the
dependence on $n$) and the fact that the module as a whole
decomposes into the direct sum of two submodules spanned by the
vectors which are odd and even in $\phi$ (the dependence on
$|m+s|_2$). As a result, \be \label{PSIR} |\Psi \rangle_\pm=
|\Psi^+ \rangle_\pm + |\Psi^- \rangle_\pm\,, \ee where, fixing
appropriately $A(n, |m+s|_2 )$,
\be \label{+d} |\Psi^+ \rangle_\pm  = \sum_{p,n=0}^\infty
\sum_{s=0,1} \f{(-1)^{p+s}}{2^{2p+s} p! (p+n+\half M+s-1)!}
(\bar{a}_a \bar{a}^a)^p \bar{\ga}^{p+n+\half M+s-1}\bar{\gb}^p
(\bar{a}_a \phi^a )^s \bar{\phi}^s
|\psi^+_n(\bar{a})\rangle_\pm\,, \ee \be \label{-d} |\Psi^-
\rangle_\pm  = \sum_{p,n=0}^\infty \sum_{s=0,1}
\f{(-1)^{p+s}}{2^{2p+s} p! (p+n+\half M+s-1)!} (\bar{a}_a
\bar{a}^a)^p \bar{\gb}^{p+n+\half M+s-1}\bar{\ga}^p (\bar{a}_a
\phi^a )^{s}
 {\phi}^s \bar{\phi}
|\psi^-_n(\bar{a})\rangle_\pm\,. \ee Although the modules
$|\Psi^\pm \rangle_\pm $ do not belong to the original Fock module
for odd $M$ because a power of the oscillator $\bar{\ga}$ or
$\bar{\gb}$ may be half-integer, they are well-defined
semi-infinite modules because one of the powers is necessarily
integer. {}From (\ref{chir}) it follows that the modules
$|\Psi^\pm \rangle_\pm $ have definite chirality
\be
\Gamma
|\Psi^{\cdots} \rangle_\pm =\pm |\Psi^{\cdots} \rangle_\pm
\ee
in agreement
with the fact that $\Gamma$ leaves invariant the space of
$osp(1,2)$ singlets.

An analogue of the formula (\ref{uf+}) is
\be
|\Psi^\pm \rangle_{\ldots} =
\bar{{\cal P}}^\pm |\psi^\pm (\bar{a} ) \rangle_{\ldots}\,,
\ee
\be
\label{ufd+} \bar{{\cal P}}^+ = \oint d \mu \exp{
(\mu^{-1}\bar{\ga} -         \mu (\f{1}{4}\bar{a}^a \bar{a}_a
\bar{\beta} +\half a_a\phi^a \bar{\phi}) )}\mu^{-\half\{a^b,
\bar{a}_b\}-2}\,, \ee \be \label{ufd-} \bar{{\cal P}}^- = \oint d
\mu \exp{(\mu^{-1}\bar{\gb} - \mu (\f{1}{4}\bar{a}^a \bar{a}_a
\bar{\ga} +
\half a_a\phi^a {\phi}))}\mu^{-\half\{a^b,
\bar{a}_b\}-2} \bar{\phi}\,. \ee Note that the additional factor
of $\bar{\phi}$ in the expression for $\bar{{\cal P}}^-$ maps the
states $|\psi^\pm (\bar{a} ) \rangle_\pm $ annihilated by $\phi$
to those annihilated by $\bar{\phi}$.

The $AdS_{M+1}$ energy operator is \be E =  \half (\bar{\ga}\ga -
\bar{\gb}\gb -[\phi , \bar{\phi} ])\,. \ee As a result, the
modules $|\Psi^\pm \rangle_\pm $ have energies \be E^+_n =
n+\half( M-1) \ee and \be E^-_n =  - n-\half (M -1)\,, \ee
respectively, depending on the upper sign $+$ or $-$ on $\Psi$.
The lowest energy of $|\Psi^+ \rangle_\pm $ therefore is \be
E^+_0= \half (M-1)\,
\ee
that is the correct value for the
conformal spinor in $M$ dimensions equal to its canonical
scaling dimension.

The $o(M,2)$ invariant norm  is defined  by the same formulas
(\ref{bpm})-(\ref{inf}). The resulting expression \be \lla
\bar{\Psi}_1^\pm |\Psi_2^\pm \rr = \sum_{n=0}^\infty 2^{n+\half M
-2}\f{1}{(n+\half M+1)!} \langle \bar{\psi}_n^\pm (a)
|{\psi}_n^\pm (\bar{a})\rangle \,,
\ee
is manifestly
positive-definite. Thus the modules $|\Psi^\pm\rangle_\pm$ are
unitary.

The noncompact generators are
\be
\label{tpmd}T^{+a} =
-i( \gb a^a - \bar{\ga} \bar{a}^a +\phi \phi^a) \,,\qquad
T^{-a} = i
( \bar{\gb} \bar{a}^a - {\ga} {a}^a -\bar{\phi} \bar{\phi}^a )\,.
\ee
The action of the generators $T^{-a}$ and $E$ has
the form
\be
\label{dit-}
T^{-a} |\Psi^+ (\psi (\bar{a}) )\rangle_\pm = |\Psi^+ \left (
i\f{\partial}{\partial \bar{a}_a } \psi (\bar{a}) \right )\rangle_\pm\,,
\ee
\be
\label{die}
E |\Psi^+ (\psi (\bar{a}))\rangle_\pm =
|\Psi^+ \left( (\bar{a}_a \f{\partial}{\partial \bar{a}_a} +\half (M -1) )
\psi( \bar{a}) \right)\rangle_\pm\,.
\ee

\section{Generalized Flato-Fronsdal theorem }
\label{FF}
An important observation  by  Flato and Fronsdal
\cite{FF} was that the tensor product of a pair of $AdS_4$
Dirac singletons \cite{Dir}
identified with the $3d$ massless particles gives rise to all $AdS_4$
massless representations. Here we extend this result to
any dimension.

Let us start with the analysis of the scalar case of $|Rac\rangle$.
The tensor product can be obtained by virtue of
doubling of oscillators $Y^A_{i} \to Y^A_{i\,1}, Y^A_{i\,2}$
with the vacuum $|0\rangle$  satisfying
\be
a^a_{1,2} |0\rangle =0\,,\qquad
\ga_{1,2} |0\rangle =0\,,\qquad
\gb_{1,2} |0\rangle =0\,.
\ee
 The tensor product of the positive energy modules
$|Rac\rangle^+\otimes |Rac\rangle^+$ is spanned by the states of the
form
\bee
\label{+2}
|\Psi^+\rangle &=&
\sum_{p,q,m,n}
\f{1}{2^{2(p+q)} q! p!(p+n+\half M-1)!(q+m+\half M-1)! }\nn\\
&{}&\ls (\bar{a_1}_l \bar{a_1}^l)^p (\bar{a_2}_k \bar{a_2}^k)^q
\bar{\ga_1}^{p+n+\half M-1}\bar{\gb_1}^p
\bar{\ga_2}^{q+m+\half M-1}\bar{\gb_2}^q
\psi^+_{nm}(\bar{a}_1 , \bar{a}_2) |0\rangle\,,
\eee
where
\bee
\psi^+_{pq}(\bar{a}_1 , \bar{a}_2)&=&
\psi^\pm_{m_1\ldots m_p \,n_1\ldots n_q}\bar{a}_1^{m_1} \ldots
\bar{a}_1^{m_p} \bar{a}_2^{n_1} \ldots
\bar{a}_2^{n_q}\,,\nn\\
&{}&\ls\ls\ls\ls\ls
\psi^\pm_{m_1\ldots m_p \,n_1\ldots n_q}\eta^{m_1 m_2} =0\,,\qquad
\psi^\pm_{m_1\ldots m_p \,n_1\ldots n_q}\eta^{n_1 n_2} =0\,.
\eee
$|\Psi^+\rangle$ satisfies
\be
t_{1\,ij}|\Psi^+\rangle =t_{2\,ij}|\Psi^+\rangle =0\,.
\ee
$|Rac\rangle^+\otimes |Rac\rangle^+$
forms a bounded energy unitary module of the HS algebra
$\huo$.

By virtue of (\ref{t-}) one observes that the lowest energy states
annihilated by $T^{-}$ are $|\Psi^+ (\psi (\bar{a}_1,\bar{a}_2
)\rangle$ with $\psi (\bar{a}_1,\bar{a}_2 )$ satisfying \be
(\f{\p}{\p \bar{a}^a_1} + \f{\p}{\p \bar{a}^a_2} ) \psi (\bar{a}_1
, \bar{a}_2) =0\,, \ee i.e.,  those with \be \psi (\bar{a}_1 ,
\bar{a}_2) = \psi_0 (\bar{a}_1 - \bar{a}_2)\,, \ee
where $\psi_0
(\bar{a})$ is an arbitrary harmonic polynomial. Using (\ref{e})
one finds that the  lowest energies are \be \label{en} E_0 = s+
M-2\,, \ee where $s$ is a degree of the polynomial $\psi_0
(\bar{a})$. Therefore
\be
\label{SS} |Rac\rangle
\otimes |Rac\rangle =\sum_{s=0}^\infty \oplus \H  (s+M-2,s,0,0\ldots)\,.
\ee
According to (\ref{mass}), the right hand side of this formula
describes for $M>2$
the direct sum of all totally symmetric massless spin
$s$ representations of the $AdS_{M+1}$ algebra $o(M,2)$.
As a result, the
tensor product of the massless scalar representation of the
conformal group in $d-1$ dimensions with $d>3$ is shown to
contain all integer spin
totally symmetric massless states in $AdS_d$, that extends the
result of Flato and Fronsdal \cite{FF} to any dimension. This
spectrum of spins exactly corresponds to that of the model of
\cite{d} that proves that the HS symmetry $\hu$ of
\cite{d} admits a unitary representation with the necessary
spin spectrum, thus satisfying the admissibility condition.

The following comments are now in order.

The spin zero field in the
$AdS_d$ HS multiplet has energy $d-3$ which is different from
the energy of conformal scalar $\half d -1$ beyond the case of $d=4$.
This is not occasional because totally symmetric massless fields
are not conformal for $d\neq 4$. It is therefore debatable whether or
not one should call this scalar field massless. We will call it
symmetrically massless
scalar to emphasize that it belongs to the HS multiplet
of symmetric massless fields and can be thought of as described by
a degenerate zero-length one row Young tableau. Other ``massless"
scalar fields with energies $d-2-p$ can be thought of as degenerate cases
of mixed symmetry gauge fields associated with $o(d-1)$-modules
described by Young tableaux with $p < \half d$ cells in the shortest
column. Conformal fields are those with the highest possible $p=\half d -1$
($d$ is even).

The case of $M=2$ (i.e., bulk $AdS_3$) is special\footnote{I am
grateful to Aleksandr Gorsky for stimulating discussion of $AdS_3$
singletons.} because the right hand side of (\ref{SS}) contains
representations corresponding to $3d$ singletons, i.e. $2d$ massless
fields of all integer spins. Indeed, in accordance with the
discussion of subsection \ref{Anti-de Sitter algebra}, the lowest
energy modules of $o(2,2)$ with the vacuum space being a
$o(2)$--module described by  Young tableaux of height 1 are $2d$
conformal fields. Thus, the bilinear tensor product of $2d$
conformal scalars gives all integer spin $2d$ conformal fields. Note
that this fact fits the admissibility condition because the $3d$ HS
gauge field dynamics is of Chern-Simons type \cite{bl} so that HS
gauge fields describe no bulk degrees of freedom analogously to the
case of $3d$ gravity \cite{3dgr}. The obtained group-theoretical
result indicates however that topological $3d$ HS interactions
should have some dynamically nontrivial boundary manifestation in
terms of $2d$ massless fields of all spins. It would interesting to
work out a dynamical realization of this phenomenon.

Let the singleton module be endowed with a Chan-Paton index
$|Rac\rangle\to |Rac\rangle^u$, $u = 1 \ldots n$. One can single out the
symmetric and antisymmetric parts $(|Rac\rangle^u
 \otimes |Rac\rangle^v)_S$ and
($|Rac\rangle^u \otimes |Rac\rangle^v)_A$ of the tensor product
$|Rac\rangle^u \otimes |Rac\rangle^v$. Since a permutation of the
tensor factors exchanges both the oscillators $\bar{a}_1$ and
$\bar{a}_2$ and the Chan-Paton indices, it follows that even (odd)
spins in $(|Rac\rangle^u \otimes |Rac\rangle^v)_S$  and
$(|Rac\rangle^u \otimes |Rac\rangle^v)_A$ are, respectively,
symmetric (antisymmetric) and antisymmetric (symmetric) in the
Chan-Paton indices. This pattern exactly corresponds to that of the
HS gauge theories of \cite{d} based on the HS gauge algebras
$ho(n,0|2\!\!:\!\![M,2])$ and $husp(n,0|2\!\!:\!\![M,2])$,
respectively. The massless states in the unsymmetrized tensor
product $|Rac\rangle^u \otimes |Rac\rangle^v$ correspond to the HS
gauge theory based on the HS gauge algebra
$hu(n,0|2\!\!:\!\![M,2])$.

Analogously one can consider higher rank tensor products of
$|Rac\rangle$. In the rank $k$ tensor product, the lowest energy
states are described by various polynomials $\psi^+(\bar{a}_1
,\ldots , \bar{a}_k)$ which are ``translationally invariant''
$\sum_i \f{\p}{\p \bar{a}_i}\psi^+(\bar{a}_1 ,\ldots , \bar{a}_k)=0$
and harmonic with respect to each variable $\bar{a}_i$. For a degree
$p$ polynomial in a rank $k$ tensor product, the $AdS_{d+1}$ lowest
energy is $E_0 = p +\half k (d-2)$. Comparing this formula with the
lowest energies for massless fields in $AdS_{d+1}$
 one finds that all states in the rank $k>2$
tensor products of $|Rac\rangle$ are massive.

To analyze  $|Di\rangle_\pm \otimes |Rac \rangle$ one observes
by virtue of (\ref{t-}), (\ref{e}),
(\ref{dit-}) and  (\ref{die}) that the lowest energy
states in  $|Di\rangle_\pm \otimes |Rac \rangle$ are described by
$\gga$-transverse (and, therefore, harmonic) tensor-spinors
$\psi (\bar{a}_1 -\bar{a}_2)_{\pm \ga}$.
As a result,
\be
\label{DiS}
 |Di\rangle_\pm \otimes |Rac\rangle=
\sum_{s=1/2,3/2\ldots } \oplus \H(s+M-2,s,\half,\half,\ldots )_\pm\,,
\ee
where $n=s-\half$ is a homogeneity degree of
$\psi (\bar{a}_1 -\bar{a}_2)_{\pm \ga}$. The
right hand side of (\ref{DiS}) contains all
totally symmetric half-integer spin massless representations
of $AdS_{M+1}$. This extends the corresponding $AdS_4$ result
of Flato and Fronsdal to any dimension $d>3$. In the special
case of $M=2$ the right hand side of (\ref{DiS}) contains
all $2d$ conformal fields of half-integer spins.

Let us now analyze a pattern of the tensor product of
two conformal spinor representations. It is convenient to analyze
$|Di\rangle_\rho \otimes{}_\kappa \langle Di |$ where
$\rho$ and $\kappa$ are the chirality factors
\be
{}_\kappa \langle \Psi^\pm |\Gamma = \kappa
{}_\kappa\langle\Psi^\pm |\,,\qquad
\Gamma |\Psi^\pm \rangle_\rho =\rho |\Psi^\pm \rangle_\rho \,,\qquad
\rho^2 = \kappa^2 = 1\,.
\ee
By virtue of
(\ref{dit-}) and  (\ref{die}) one finds that the lowest energy
states in $|Di\rangle_\rho \otimes{}_\kappa \langle Di |$
  are described by harmonic tensor bispinors
$\psi (\bar{a}_1 +\bar{a}_2)_{\rho \kappa}{}_{  \ga}{}^\gb$,
which are  $\gamma$-transverse both in $\ga$ and in $\gb$.
Equivalently, one can write
\be
\psi ({\bar{a}})_{\rho \kappa}{}_{  \ga}{}^\gb =\sum_{m=0}^M
\sum_{n=0}^\infty
\psi_{\rho\kappa}^{[b_1 \ldots b_m]|\{a_1\ldots a_n \}}
\gga_{[b_1\ldots b_m]}{}_\ga{}^\gb\, \bar{a}_{a_1}\ldots \bar{a}_{a_n}\,,
\ee
where $\gga_{[b_1\ldots b_m]}$ are totally antisymmetrized
products of $\gamma$ matrices.
$\psi_{\rho\kappa}^{[b_1 \ldots b_m]\{a_1\ldots a_n \}}$
is totally symmetric in the indices $a$ and totally antisymmetric
in the indices $b$. It is easy to see that
the left and right $\gamma$-transversality imply that
$\psi_{\rho\kappa}^{[b_1 \ldots b_m]\{a_1\ldots a_n \}}$
is traceless and that antisymmetrization over any $m+1$
indices must give zero. In other words it is described by the
following traceless Young tableau of $o(M)$
\bee
\begin{picture}(40,50)
{
\put(00,35){\line(1,0){40}}%
\put(00,30){\line(1,0){40}}%
\put(00,25){\line(1,0){05}}%
\put(00,20){\line(1,0){05}}%
\put(00,15){\line(1,0){05}}%
\put(00,10){\line(1,0){05}}%
\put(00,05){\line(1,0){05}}%
\put(00,00){\line(1,0){05}}%
\put(00,00){\line(0,1){35}} \put(05,00.0){\line(0,1){35}}
\put(10,30.0){\line(0,1){05}}
\put(15,30.0){\line(0,1){05}} \put(20,30.0){\line(0,1){05}}
\put(25,30.0){\line(0,1){05}} \put(30,30.0){\line(0,1){05}}
\put(35,30.0){\line(0,1){05}} \put(40,30.){\line(0,1){05}}
}
\put(12,37.){\scriptsize  $n+1$}
\put(-10,20){\scriptsize  $m$}
\end{picture}.
\label{ddd}
\eee
According to (\ref{die}) the energies of these states are
\bee
E_0 &=&n +M-1= s+M-2\,, \qquad s>0,\quad m<M  \nn\\
E_0 &=& M-1 \,,\qquad\qquad\qquad\qquad s=0\,,\quad or\quad s=1, \quad m=M\,,
\eee
where $s$ is the length of the upper row of the tableau
(\ref{ddd}). This means that all states with $s>0$
in the tensor  product are massless except for
those with $s=1, M> m>0$ and $s=0$
\bee
\label{didir}
|Di\rangle \otimes \langle Di |
&=&
2 \H(M-1,0,0\ldots) \oplus
\sum_{s=1}^\infty \oplus \Big (2 \sum_{m=0}^{[\f{M}{2}]} \H(s+M-2,s,
\underbrace{1,1\ldots 1}_{m},0,0\ldots)
\,\nn\\
&{}&\ls\oplus \H(s+M-2,s, \underbrace{1,1\ldots 1}_{M/2-1})
\oplus \H(s+M-2,s, \underbrace{1,1\ldots
1}_{M/2-2},-1) \Big )\,,
\eee
where the last two
terms appear only for even $M$ when the (anti)selfduality
condition can be imposed. The appearance of
massive totally antisymmetric  fields\hfil
\\ $\H(M-1,1,1,\ldots , 0,0\ldots)$
(which, however, become massless in the flat limit)
in the higher dimensional HS multiplets
is analogous to the case of the spin 0 field in
$AdS_4$ \cite{FF}.

The formula (\ref{didir}) is true for Dirac vacuum spinors.
Imposing the chirality conditions for even $M$ we have the
type A situation with the opposite chiralities
 and type B case with the same chiralities.
In the type A case the column in (\ref{ddd})
contains an odd number of cells while in the type
B situation the column in (\ref{ddd})
contains an even numbers of  cells. In addition, the tableaux
(\ref{ddd}) with $m$
cells in a column are equivalent (dual) to those with $M-m$
cells. In particular, in the type B case the representation
with $m=M/2$ is selfdual or antiselfdual depending on the
chirality $\rho$ of  $|Di\rangle_\rho$. For the special case
of $M=2$ we obtain all totally symmetric integer spin $s>0$
$2d$ conformal fields in the type $A$ case. For the Dirac or
type $B$ $M=2$ case, the tensor product of two $2d$ conformal
spinor modules contains a $3d$ massive scalar field with
$E_0 =1$. It would be interesting to see what is a field-theoretical
realization of this system.

For odd $M$ the operator $\Gamma$ is central. As a result, only
the type B case is nontrivial for odd $M$. Since the
corresponding tableaux (\ref{ddd}) are
selfdual, the resulting expansion contains
all inequivalent representations in a single copy.

Using the results of \cite{ASV}, we conclude that the list of gauge
fields resulting from gauging of $\su$ just matches the list of
massless states in the tensor product $(|Rac\rangle\oplus|Di\rangle
) \otimes (\langle Rac| \oplus \langle Di |)$. Thus, the
superalgebra $\su$ and its chiral versions satisfy the admissibility
condition and therefore are expected to give rise to consistent
supersymmetric HS gauge theories in any dimension with totally
symmetric fermionic massless fields of all half-integer spins.
Leaving details of the exact formulation for a future publication
let us mention that the form of the nonlinear dynamical equations
for the supersymmetric case is essentially the same as that of
\cite{d} for the purely bosonic case modulo extension of the
generating elements of the algebra with the Clifford fermions
$\phi^A$ and spinor generating elements $\chi_\mu$ and $ \bar{\chi}
^\mu $. The same is true for the algebras $\hunm$ with the
nontrivial (spin 1) Yang-Mills algebra $u(n)\oplus u(m)$ and their
orthogonal and symplectic reductions.

It is worth to note that although  HS theories
based on the superalgebra\\ $\su$ are supersymmetric
in the HS sense, they are not necessarily supersymmetric in the
standard sense. As explained in section \ref{Spinorial realization},
$\su$ contains usual $AdS$ superalgebras as subalgebras only for
some lower $M$.

\section{Unfolded equations for conformal fields}
\label{Unfolded equations for conformal fields}
As pointed out in  \cite{Gun1}, there is a duality between
unitary modules of single-particle quantum states and nonunitary modules
underlying classical field equations. In
\cite{Shaynkman:2001ip,BHS}
it was shown for the  $3d$ and $4d$ conformal systems
 that the corresponding duality has a form of certain (nonunitary)
Bogolyubov transform. Let us show that the same
is true in any dimension by deriving unfolded form of free
conformal massless equations in $M$ dimensions.

Let us introduce the following basis of oscillators:
 $y^\pm = {Y_1^{M}\pm Y^{M+1}_1}$,
$p^\pm = \frac{i}{2} ({Y_2^{M}\pm Y_2^{M+1}})$ and
$y^n=Y_1^n$,
$p^n= \frac{i}{2}Y_2^n$ with $n=0\ldots M-1$ being
$M$-dimensional Lorentz indices. The nonzero
commutation relations are
\be
\label{lor}
[y^n , p^m ] =- \eta^{nm}\,,\qquad
[y^\pm  , p^\pm ] =0\,,\qquad [y^\pm  , p^\mp ] =2\,,
\ee
where $\eta^{nm}$ is the Minkowski metric with the signature
$(1,M-1)$. Now we introduce a non-unitary  Fock module $F$
with the vacuum state
\be
p^n |0\rangle =0\,,\qquad y^\pm  |0\rangle =0\,.
\ee
Its general element is
\be
|\Phi \rangle = \phi (y^n,p^+,p^- ) |0\rangle \,.
\ee

The submodule $SF$ to be associated with
the classical scalar field in $M$ dimensions is spanned by the
$sp(2)$ invariant states satisfying $t_{ij}|\Phi \rangle=0$.
To describe a massless scalar field
in the Minkowski space $R^{M}$
we consider sections of the trivial fiber bundle $R^{M}\times SF$
\be
|\Phi (x) \rangle = \phi (y^n,p^+,p^- |x) |0\rangle \,.
\ee
The conditions  $t_{ij}|\Phi \rangle=0$ imply that
$\phi (y^n,p^+,p^- |x)$ has a form analogous to (\ref{+}) and
(\ref{-})
\be
\phi (y,p^\pm|x ) =
\phi^+ (y,p^\pm |x )+
\phi^- (y,p^\pm|x ) \,
\ee
with
\be
\label{+l}
\phi^+ (y,p^\pm|x  ) =
\sum_{p,n} \f{1}{2^{2p} p! (p+n+\half M-1)!}
(y^m y_m )^p
(p^+)^{p+n+\half M-1}(p^-)^p \phi^+_n(y|x)\,,
\ee
\be
\label{-l}
\phi^- (y,p^\pm|x   ) =
\sum_{p,n} \f{1}{2^{2p} p! (p+n+\half M-1)!}
(y^m y_m )^p
(p^-)^{p+n+\half M-1}(p^+)^p \phi^+_n(y|x)
\ee
and $\phi^\pm_n(y|x)$ are arbitrary degree $n$ harmonic polynomials
of $y$, i.e.
\be
\label{tr0}
\phi^\pm_n(y|x)=\phi^\pm_{{m_1}\ldots {m_n}}y^{m_1} \ldots
y^{m_n} \,,\qquad \psi^\pm_{kl m_3\ldots m_n}\eta^{kl} =0\,.
\ee

To unfold some dynamical equations means to
reformulate them in the form of appropriate
zero curvature and covariant constancy conditions
(for more details we refer the reader to the original
paper \cite{Ann} and to \cite{gol,sah3}).
In particular, the equations for matter fields
and massless fields
reformulated in terms of covariant curvatures  (like, for example,
 Maxwell equations in terms of field strengths) have a form of
covariant constancy  equations on certain 0-forms called Weyl
0-forms. This name is borrowed from gravity where the corresponding
covariant constancy  equations describe differential restrictions on
the Weyl tensor and all its derivatives. The Weyl 0-forms are
sections of the fiber bundle over space-time with the fiber space
dual to the space of single-particle quantum states by a Bogolyubov
transform.

In our case the unfolded equations are
\be
\label{unfsc}
D |\Phi^+(x) \rangle =0\,,
\ee
 where $D= d + \go_0$ is the
covariant derivative with a flat connection $\go_0\,,$
\be
D^2 =0\,,
\ee
which takes values in the conformal algebra $o(M,2)$
acting on the fiber module $SF$. To describe conformal
field equations in flat (i.e., Minkowski) space-time
 one chooses $\go_0$ to take values in the
Poincare subalgebra of the conformal algebra. To use Cartesian
coordinates, one takes the connection $\go_0$ in the form $\go_0 =
dx^n P_n$, where $P_n$ are generators of translations of the
Poincare algebra. In our case, $P_n =  y^- p_n$ and the equation
(\ref{unfsc}) gets the form \be \label{unf} dx^n (\f{\p}{\p x^n }
+ y^- p_n ) |\Phi^+ (x) \rangle =0\,. \ee In terms of components
$\phi^+_n(y|x)$ it is equivalent to the infinite chain of
equations
\be
dx^m \left (\f{\p}{\p x^m }\phi^+_n(y|x) + \f{\p}{\p
y^m } \phi^+_{n+1}(y|x)\right )=0 \,
\ee
which  expresses all
higher components $\phi^+_n(y|x)$ with $n>0$ via higher
$x$-derivatives  of $\phi^+_0(x)$ identified with the physical
scalar field which satisfies the Klein-Gordon equation as a result
of the conditions (\ref{tr0}). The case of a scalar field in any
dimension was considered in detail in \cite{SHV}. Let us note
that from the unfolded form of the massless scalar field equation
interpreted as a covariant constancy condition it immediately
follows (see e.g. \cite{sah3}) that the massless scalar field
equation is invariant under the global symmetry algebra
 $\hu$ that provides an
elementary proof of the result obtained by Eastwood
\cite{East}.

Unfolded form of the fermionic massless equations is obtained
analogously by using the
spinorial module $|\Phi\rangle_\nu$ and the realization (\ref{TS})
of the conformal generators. The resulting equations can be found
in \cite{STV} where the unfolded reformulation of all possible
conformal field equations was given.

\section{Conclusion}
\label{Conclusion} It is shown that $AdS_d$ HS global symmetry
algebras underlying HS gauge theories of totally symmetric
massless fields in $AdS_d$ of \cite{d} admit unitary
representations with the spectra of states matching those of the
respective field-theoretical HS models. The states of the $AdS_d$ HS
models of \cite{d} correspond to the tensor product of the
singleton modules identified with the space of single-particle states
of the conformal scalar field in $d-1$ dimension. This fact extends
the original observation of Flato and Fronsdal for $AdS_4$ \cite{FF}
to any dimension and provides a group-theoretical basis for the $AdS/CFT$
correspondence between  conformal boundary models and $AdS_d$ bulk
HS theories. The group-theoretical analysis of this paper fits the
field-theoretical analysis of the $AdS/CFT$ correspondence
between bulk HS models and boundary conformal models of scalar fields
carried out in \cite{Das:2003vw,Gopakumar:2003ns},
 based on the observation that
conserved currents built of a massless scalar in $d$ dimensions
match the list of  on-mass-shell HS gauge fields in
the $d+1$ dimensional bulk.
In particular, the bilocal field
introduced in \cite{Das:2003vw} is a field-theoretical realization of
the tensor product of a pair of singletons.

The extension to the supersymmetric case gives rise
to HS superalgebras acting on the boundary  conformal
scalar and spinor as well as
on infinite sets of totally symmetric massless bulk bosons
and fermions  and mixed symmetry massless fields described by
hook tableaux with one row and one column.
 We argue that the bulk HS theories
associated with the conformal spinor fields on the boundary are
described by the nonlinear field equations having essentially the
same form as that of \cite{d} for totally symmetric massless
fields. An interesting project for the future is to investigate
whether there exists a generalization of the obtained results to a
broader class of HS gauge fields in the bulk, which correspond to
massless representations  of a generic mixed symmetry
type in $AdS_d$. {}From the field-theoretical side this requires further
study of the mixed symmetry gauge fields in any dimension because,
despite considerable progress achieved in the literature
\cite{mixed,ASV}, the full covariant formulation in $AdS_d$ is
still lacking even at the free field level for generic $d$. Note
that the full formulation of free totally symmetric massless
fields in $AdS_d$ was obtained in \cite{LV,vasfer} in terms of HS
gauge connections and in \cite{BPT,ST} by using the BRST
formalism. Also it is worth to mention that the structures of
generic massless mixed symmetry fields in flat space and $AdS_d$
are essentially different: an irreducible massless field in $AdS_d$
decomposes into a family of massless fields in the flat limit
\cite{BMV}.

Finally, let us mention that the case of $AdS_3$ singletons is
special because bilinear tensor products of $2d$ conformal
scalar and spinor contain
infinite sets of $d=2$ HS fields rather than $d+1$ fields as
it happens for all $d>2$.
This fact agrees with the field-theoretical description because
HS gauge field dynamics in $AdS_3$ is of Chern-Simons type \cite{bl}
so that HS gauge fields describe no bulk degrees of freedom. The
obtained group-theoretical
result indicates however that topological $3d$ HS interactions should
have some dynamically nontrivial boundary manifestation in terms of
$2d$ massless fields of all spins.

\section*{Acknowledgments}
The author is grateful to Per Sundell and Aleksandr Gorsky
for stimulating discussions
and to Laurent Baulieu for hospitality at LPTHE, Paris U. VI-VII
in the fall of 2003 where most of this work was done.
This research was supported in part by INTAS, Grant No.00-01-254,
 and the RFBR, Grant No.02-02-17067.

\section*{Appendix. Young tableaux}
\label{Appendix}

For the reader's convenience we summarize here some elementary
properties of Young tableaux.

Let a $sl_M$ tensor $A^{a^1_1\ldots a^1_{m_1},a^2_1\ldots a^2_{m_2},
a^3_1\ldots a^3_{m_3} \ldots a^p_1\ldots a^p_{m_p}}$ be symmetric
in the indices
$a^i_k$ for any fixed $i$. It corresponds to the representation
of $sl_M$ described by the Young tableau $Y(m_1, m_2 ,\ldots m_p)$
\sbox{\toch}{\circle*{1}}
\sbox{\gorp}{\line(1,0){5}}
\sbox{\verp}{\line(0,1){5}}
\bee   
\begin{picture}(85,40)(0,5)%
{
\put(0,10){\line(0,1){21}}
\multiput(0,5)(5,0){2}{\usebox{\gorp}}%
\multiput(0,5)(5,0){3}{\usebox{\verp}}%
\put(15,5){\tiny $m_{p}$}
\multiput(0,10)(5,0){6}{\usebox{\gorp}}%
\multiput(0,10)(5,0){7}{\usebox{\verp}}%
\put(36,10){\tiny $m_{p-1}$}
\multiput(0,15)(5,0){6}{\usebox{\gorp}}%
\multiput(3,17.5)(5,0){7}{\usebox{\toch}}%
\multiput(3,20)(5,0){7}{\usebox{\toch}}%
\multiput(3,22.5)(5,0){8}{\usebox{\toch}}%
\multiput(3,25)(5,0){8}{\usebox{\toch}}%
\multiput(3,27.5)(5,0){9}{\usebox{\toch}}%
\multiput(0,30)(5,0){10}{\usebox{\gorp}}%
\multiput(0,30)(5,0){11}{\usebox{\verp}}%
\put(56,30){\tiny $m_2 $}
\multiput(0,35)(5,0){12}{\usebox{\gorp}}%
\multiput(0,35)(5,0){13}{\usebox{\verp}}%
\put(65,35){\tiny $m_1$}
\multiput(0,40)(5,0){12}{\usebox{\gorp}}}
\end{picture}
\eee
if the tensor $A$ is such that, for any fixed  $i$,
total symmetrization of all indices
$a_k^i$ with any index $a^l_j$ such that $l>i$ gives zero.
Let $y_a^i$ be auxiliary variables with $a=1,2\ldots M$,
$i=1,2\ldots p$. The tensors
$A^{a^1_1\ldots a^1_{m_1},a^2_1\ldots a^2_{m_2},
 \ldots a^p_1\ldots a^p_{m_p}}$ can be identified with the
coefficients of the polynomials
\be
\label{dec}
A(y) = \sum_{m_1 , m_2 , \ldots}^\infty
A^{a^1_1\ldots a^1_{m_1},a^2_1\ldots a^2_{m_2}, \ldots a^p_1\ldots a^p_{m_p}}
y_{a^1_1}^1  \ldots y_{a^1_{m_1}}^1
y_{a^2_1}^2  \ldots y_{a^2_{m_2}}^2
 \ldots
y_{a^p_1}^p  \ldots y_{a^p_{m_p}}^p \,.
\ee
Note that the polynomials $A(y)$ can be written as
\be
A(y) = \sum_{n=0}^\infty A^{a_1\ldots a_n}_{i_1\ldots i_n}
y_{a_1}^{i_1}\ldots y_{a_n}^{i_n}\,.
\ee
In these terms, the Young conditions take the simple form
\be
\label{hw}
y^i_a \f{\p}{\p y^j_a} A(y) =0  \qquad i < j \,,
\ee
\be
\label{cart}
y^i_a \f{\p}{\p y^i_a} A(y) = m_i P(y)\qquad \mbox{ no summation over  $i$}\,.
\ee

The operators
\be
t_a{}^b = y_a^i \f{\p}{\p y^i_b}
\ee
and
\be
\label{l}
l^i{}_j = y_a^i \f{\p}{\p y^j_a}
\ee
form the algebras $gl_M$ and $gl_p$, respectively.
They are mutually commuting and are called Howe dual.
It is useful to observe that the conditions (\ref{hw})
are the highest weight conditions for the algebra
$sl_p\subset gl_p$.
The $gl_M$ invariant conditions
(\ref{cart}) fix some (integral) highest weight of
$sl_p$ together with an eigenvalue of the central element of
$gl_p$. Rectangular Young tableaux
\sbox{\toch}{\circle*{1}}
\sbox{\gor}{\line(1,0){5}}
\sbox{\ver}{\line(0,1){5}}
\bee   
\label{block}
\begin{picture}(150,60)(0,65)%
{
\multiput(0,70)(5,0){19}{\usebox{\gor}}%
\multiput(0,70)(5,0){20}{\usebox{\ver}}%
\multiput(0,75)(5,0){19}{\usebox{\gor}}%
\multiput(0,75)(5,0){20}{\usebox{\ver}}%
\multiput(0,80)(5,0){19}{\usebox{\gor}}%
\multiput(0,80)(5,0){20}{\usebox{\ver}}%
\multiput(0,85)(5,0){19}{\usebox{\gor}}%
\multiput(0,85)(5,0){20}{\usebox{\ver}}%
\multiput(0,90)(5,0){19}{\usebox{\gor}}%
\multiput(0,90)(5,0){20}{\usebox{\ver}}%
\multiput(0,95)(5,0){19}{\usebox{\gor}}%
\multiput(0,95)(5,0){20}{\usebox{\ver}}%
\multiput(0,100)(5,0){19}{\usebox{\gor}}%
\multiput(0,100)(5,0){20}{\usebox{\ver}}%
\multiput(0,105)(5,0){19}{\usebox{\gor}}%
\multiput(0,105)(5,0){20}{\usebox{\ver}}%
\multiput(0,110)(5,0){19}{\usebox{\gor}}%
\multiput(0,110)(5,0){20}{\usebox{\ver}}%
\multiput(0,115)(5,0){19}{\usebox{\gor}}%
\multiput(0,115)(5,0){20}{\usebox{\ver}}%
\multiput(0,120)(5,0){19}{\usebox{\gor}}%
\put(50,123){$m$}
\put(105,90){$p$}
}
\end{picture}
\eee
with $m_i = m$, which we call blocks,
have a special property that they are $sl_p$ singlets
\be
\label{bl}
y^i_a \f{\p}{\p y^j_a} P(y) = \f{1}{p}\delta^i_j
y^k_a \f{\p}{\p y^k_a} P(y) \,.
\ee
(Note that (\ref{bl}) is a consequence of (\ref{hw}), (\ref{cart})
with $m_i= const$ along with the fact that the representations are
finite dimensional because $A(y)$ is a polynomial. Combinatorial
proof of this fact in terms of components of tensors is
also elementary).

{}From the definition of the Young tableau
$Y(m_1 , m_2 ,\ldots m_p)$
it follows that $m_1 \geq m_2 \geq m_3 \ldots$
(otherwise  the corresponding tensors are zero).
For the same
space we will also use notation with square brackets
${Y}[l_1 , l_2 ,\ldots ]$
where $l_1 , l_2 ,\ldots$ are heights of columns
\sbox{\verp}{\line(1,0){5}}
\sbox{\gorp}{\line(0,1){5}}
\bee   
\begin{picture}(55,60)(0,-15)%
{
\put(10,45){\line(1,0){20}}
\multiput(40,40)(0,-5){2}{\usebox{\gorp}}%
\multiput(35,45)(0,-5){3}{\usebox{\verp}}%
\put(5,-21){\tiny $l_{1}$}
\multiput(035,40)(0,-5){6}{\usebox{\gorp}}%
\multiput(030,45)(0,-5){7}{\usebox{\verp}}%
\put(11,-6){\tiny $l_{2}$}
\multiput(030,40)(0,-5){6}{\usebox{\gorp}}
\multiput(27.5,42)(0,-5){7}{\usebox{\toch}}%
\multiput(25,42)(0,-5){7}{\usebox{\toch}}%
\multiput(22.5,42)(0,-5){8}{\usebox{\toch}}%
\multiput(20,42)(0,-5){8}{\usebox{\toch}}%
\multiput(17.5,42)(0,-5){9}{\usebox{\toch}}%
\multiput(015,40)(0,-5){9}{\usebox{\gorp}}%
\multiput(010,45)(0,-5){10}{\usebox{\verp}}%
\put(31,09){\tiny $l_{q-1}$}
\put(36,29){\tiny $l_{q}$}
\multiput(10,40)(0,-5){12}{\usebox{\gorp}}%
\multiput(05,45)(0,-5){13}{\usebox{\verp}}%
\multiput(05,40)(0,-5){12}{\usebox{\gorp}}}
\end{picture}
\eee
Obviously, one has
$l_1 \geq l_2 \geq l_3 \ldots$
 and $l_1 \leq M$ (because antisymmetrization over any $M+1$ indices
$a$ taking $M$ values gives zero).

The realization of
${Y}[l_1 , l_2 ,\ldots l_q]$ with manifest antisymmetrization is
achieved in terms of polynomials  $F(\phi^\ga_a)$ of fermions
\be
\phi^\ga_a \phi^\gb_b =-\phi^\gb_b  \phi^\ga_a  \,,\qquad
\ga, \gb = 1\ldots q\,.
\ee
The Young properties equivalent to (\ref{hw}) and (\ref{cart})
are
\be
\label{hwf}
\phi^\ga_a \f{\p}{\p \phi^\gb_a} F(\phi) =0  \qquad \ga < \gb\,,
\ee
\be
\label{cartf}
\phi^\ga_a \f{\p}{\p \phi^\ga_a} F(y) = l_\ga F(y)\qquad
\mbox{ no summation over  $\ga$}\,.
\ee
The coefficients
$F^{a^1_1\ldots a^1_{l_1},a^2_1\ldots a^2_{l_2}, \ldots a^p_1\ldots a^q_{l_q}}$
of
\be
\label{decf}
F(\phi ) = \sum_{l_1 , l_2 , \ldots , l_q}^\infty
F^{a^1_1\ldots a^1_{l_1},a^2_1\ldots a^2_{l_2}, \ldots a^p_1\ldots a^q_{l_q}}
\phi_{a^1_1}^1  \ldots \phi_{a^1_{l_1}}^1
\phi_{a^2_1}^2  \ldots \phi_{a^2_{l_2}}^2
 \ldots
\phi_{a^q_1}^q  \ldots \phi_{a^q_{l_q}}^q \,
\ee
are manifestly antisymmetric in the groups of indices $a_k^\ga$ with
fixed $\ga$ associated with the $\ga^{th}$ columns of the tableau.
The condition (\ref{hwf}) requires that total antisymmetrization of all
indices associated with some column with any index from any
subsequent column gives zero. Again, the conditions (\ref{hwf}) and
(\ref{cartf}) are highest weight conditions for the algebra $gl_q$,
formed by the  operators
\be
f^\ga{}_\gb = \phi_a^\ga \f{\p}{\p \phi^\gb_a} \,,
\ee
which is Howe dual to the $gl_M$ formed by
\be
s_a{}^b = \phi_a^\ga \f{\p}{\p \phi^\ga_b}\,.
\ee
Rectangular (block) tableaux are $sl_q$ singlets.

It is not hard to see that the spaces of tensors with manifest
symmetry and antisymmetry,
${Y}(m_1 , m_2 ,\ldots m_p)$ and
${Y}[l_1 , l_2 ,\ldots l_q]$, respectively, associated with the same
Young tableau, are isomorphic.

If one is interested in representations of $o(M)$ rather than $sl_M$,
the set of conditions (\ref{hw}), (\ref{cart}) is supplemented
with
\be
\label{tr}
\eta_{ab}  \f{\p^2}{\p y^j_a\p y^i_b}          A(y) =0 \,,
\ee
where $\eta^{ab}$ is the $o(M)$ invariant metric. This is
simply the condition that all $o(M)$ tensors in the decomposition (\ref{dec})
are traceless.  For such spaces we  use notation
 $Y^{tr}(m_1 , m_2 ,\ldots m_p)$.
Note that for $o(M)$, the bosonic Howe dual algebra
extends to $sp(2p)$ generated by (\ref{l}) along with
the operators
\be
l^{ij} =    \eta^{ab}  y_a^i y^j_b          \,,\qquad
l_{ij} =\eta_{ab}  \f{\p^2}{\p y^j_a\p y^i_b} \,,
\ee
which form the standard oscillator realization of $sp(2p)$.
In the fermionic realization the equivalent condition is
\be
\eta_{ab}  \f{\p^2}{\p \phi^\ga_a\p \phi^\gb_b} F(\phi) =0 \,.
\ee
The fermionic Howe dual algebra of $o(M)$ is $o(2q)$.

An important fact proved in section
 \ref{Simplest bosonic higher spin algebras} is that
\be
\label{kons}
Y^{tr}[l_1 , l_2 ,\ldots l_q]=0\qquad \mbox{if\quad $l_1 +l_2 >M$}.
\ee
Note that this
identity is insensitive to the full Young properties. The only important
properties are the tracelessness and total
antisymmetry of each of the two groups of indices which together have
more than $M$ indices.

Traceless $o(M)$ tableaux can be dualized by contracting
the $\gep$ symbol with the first column. For even $M$,
one can define (anti)selfdual tableaux
$Y_\pm^{tr} [M/2 , l_2, l_3, \ldots , l_q ]$
with the height of the first column $M/2$.
Note that for any Young tableau there is at most one way to
define selfduality because the maximal vertical block
in the antisymmetric basis is symmetric
with respect to its column interchange.

\end{document}